\documentclass[preprints,article,accept,oneauthor,a4paper]{Definitions/mdpi}

\usepackage{quantikz}   
\usepackage{braket}     

\usepackage{lineno}
\nolinenumbers
\firstpage{1} 
\pubvolume{1}
\issuenum{1}
\articlenumber{0}
\pubyear{2026}
\copyrightyear{2026}
\externaleditor{Christoph Bandt} 
\datereceived{3 May 2026} 
\daterevised{24 June 2026} 
\dateaccepted{25 June 2026} 
\datepublished{ } 

\usepackage{xcolor}
\usepackage{soul}
\usepackage{graphicx}

\Title{Quantum Circuit Learning for Volatility Modeling: Multifractal Analysis of Realized Volatility Time Series}

\Author{{Tetsuya Takaishi} 
 \orcidA{}}

\address[1]{Department of Liberal Arts, Hiroshima University of Economics, Hiroshima 731-0192, Japan; tt-taka@hue.ac.jp\\
}

\nolinenumbers

\abstract{
Herein, we propose a quantum circuit learning  framework for modeling the
realized volatility (RV) of Bitcoin and investigate the statistical
properties of the predicted time series through multifractal analysis.
Unlike conventional GARCH-type models, which require a pre-specified
functional form for the volatility process, a parameterized quantum
circuit directly approximates the volatility function from empirical
data, eliminating the need for explicit model selection.
Using five-minute Bitcoin price data, we construct daily RV, train a
single-qubit parameterized quantum circuit, and
generate a long synthetic time series from the optimized quantum circuit.
Multifractal Detrended Fluctuation Analysis is applied to
calculate the generalized Hurst exponent $h(q)$, the singularity
spectrum $f(\alpha)$, and the multifractal scaling exponent $\tau(q)$.
The predicted return series exhibits $h(2)\approx 0.5$,
consistent with near-random dynamics, and 
both the predicted and the empirical return series display
multifractality that partially persists after random shuffling.
The increment series of RV shows pronounced
anti-persistence with $h(2)\approx 0.05$--$0.1$, consistent with the
rough volatility hypothesis.
These results demonstrate that a simple single-qubit parameterized quantum
circuit captures qualitatively some observed properties in
Bitcoin volatility dynamics.
}


\begin{document}


\section{Introduction}
Risk management is a central concern for participants in financial markets, where estimating the risk associated with asset holdings and preventing large future losses are essential tasks. Volatility is one of the most widely used measures of financial risk, and forecasting future volatility plays a crucial role in the safe and efficient management of financial assets. A common approach to volatility forecasting is to construct models that capture the dynamics of financial time series. In this context, incorporating empirical properties of financial data is a key requirement for building effective models.

Several stylized facts are known to appear universally across different classes of financial assets~\cite{Cont2001QF}. Among these, volatility clustering is one of the most prominent: periods of high volatility tend to be followed by high volatility, and periods of low volatility by low volatility. To capture this feature, Engle introduced the Autoregressive Conditional Heteroskedasticity (ARCH) model~\cite{Engle1982autoregressive}, in which volatility is modeled as an autoregressive function of past squared returns. Bollerslev later generalized this framework to the {Generalized ARCH (GARCH)} model~\cite{Bollerslev1986JOE}, which has since become a standard tool in \mbox{volatility modeling.}

Another important empirical property of asset returns, especially stock returns, is the leverage effect~\cite{Black1976,Christie1982stochastic}, which refers to the asymmetric response of volatility to positive and negative returns: volatility tends to increase more following negative returns than following positive ones. Since the standard GARCH model cannot capture such asymmetry, several extensions have been proposed, including the Exponential GARCH (EGARCH)~\cite{Nelson1991Econ}, Threshold GARCH (TGARCH)~\cite{Glosten1993JOF}, Quadratic ARCH (QARCH)~\cite{Sentana1995RES}, Rational GARCH (RGARCH)~\cite{takaishi2017rational,takaishi2018volatility}, and Asymmetric Power GARCH (APGARCH)~\cite{ding1993long} models.

A further class of models incorporating stochastic dynamics is the
Stochastic Volatility (SV) ~\cite{taylor1982,taylor1986modelling}, which augments the volatility
equation with a stochastic disturbance term. Whereas GARCH-type models
can generally be estimated via maximum likelihood, which is
computationally straightforward, maximum likelihood estimation of SV
models is intractable; Bayesian estimation (It is
 also possible to estimate the parameters of a GARCH model within a Bayesian framework. For example, see \cite{Takaishi2006/10}.)  is therefore the predominant
approach, implemented through Markov Chain Monte Carlo methods developed
specifically for this \mbox{purpose~\cite{kim1998stochastic,jacquier2002bayesian,omori2007stochastic}.}
In lattice Quantum ChromoDynamics (QCD) simulations~\cite{gupta1998introduction,lippert2007hybrid}, physical observables are typically computed using Markov chain Monte Carlo methods, among which the standard algorithm is the Hybrid Monte Carlo (HMC) method~\cite{duane1987hybrid}. Although HMC was originally developed (For studies on methods that improve the efficiency of the HMC algorithm in lattice QCD simulations, see, for example, References 
 \cite{sexton1992hamiltonian,Takaishi1997Fast,hasenbusch2001speeding,takaishi2000choice}. In realistic lattice QCD simulations, it is necessary to handle an odd number of fermion flavors, and several variants of the HMC algorithm capable of performing such simulations have been proposed, including the following~\cite{takaishi2002odd,clark2006rational,clark2007accelerating}.)  within the context of lattice QCD computations, it has since been adopted across a wide range of fields, where it is also referred to as Hamiltonian Monte Carlo. Several studies~\cite{takaishi2009bayesian,takaishi2014RSV} have applied this HMC framework to the Bayesian estimation of SV-type models.

Another stylized fact concerns the long-memory property of volatility. Although GARCH-type models imply short-memory dynamics, empirical studies have shown that volatility often exhibits long-range dependence~\cite{ding1993long,andersen2003modeling}. To address this, long-memory models such as the Autoregressive Fractionally Integrated Moving Average (ARFIMA)~\cite{granger1980introduction} and Fractionally Integrated GARCH (FIGARCH)~\cite{baillie1996fractionally,tayefi2012overview} models have been proposed.

More recently, Gatheral et al.~\cite{gatheral2018volatility} reported that the increments of realized volatility (RV) exhibit a Hurst exponent of approximately $H \approx 0.1$, implying anti-persistent behavior. This phenomenon, known as rough volatility
, has been confirmed across various asset classes~\cite{bennedsen2022decoupling,livieri2018rough,floc2022roughness,takaishi2025multifractality}, and rough fractional volatility models have been shown to provide an excellent fit to empirical data~\cite{gatheral2018volatility}.
The CBOE Volatility Index (VIX), which is computed from option prices on the S\&P 500 index, is a widely used measure of market uncertainty and stock market volatility. Studies on the Hurst exponent of the VIX time series have shown that, similar to RV, its value is less than 0.5, indicating anti-persistence, or the rough volatility property~\cite{bariviera2023disentangling,takaishi2025impact}.
Under the Mixture of Distributions Hypothesis~\cite{clark1973subordinated,tauchen1983price,andersen1996return}, price fluctuations are driven by the amount of information flowing into the market, and Clark~\cite{clark1973subordinated} employs trading volume as a proxy for information arrival. In this framework, trading volume and volatility are expected to exhibit a strong correlation (As an illustration of the strong correlation between trading volume and volatility, it has been pointed out that introducing trading volume into GARCH models leads to a reduction in the GARCH   effect~\cite{lamoureux1990heteroskedasticity}. Subsequently, a substantial body of research has incorporated volume variables into GARCH-type models. See, for example, ~\cite{sharma1996heteroscedasticity,miyakoshi2002arch,bose2015examining,takaishi2016relationship}.),  
implying that the volume time series should share statistical properties similar to those of the volatility time series. Indeed, it has been shown that the Hurst exponent of trading volume time series is also less than 0.5, reflecting the rough volatility property~\cite{takaishi2022Hurst}.

Given the wide variety of volatility models, each producing different estimates and forecasts, selecting an appropriate model for a given dataset remains a practical challenge. Model-free measures such as RV~\cite{andersen1998answering,andersen2003modeling,barndorff2002econometric,mcaleer2008realized}, constructed from high-frequency data, provide accurate estimates of daily volatility. However, forecasting still requires a model, leading to the development of hybrid approaches such as Realized GARCH~\cite{hansen2012realized,hansen2016exponential}, Realized \mbox{SV~\cite{takahashi2009estimating,koopman2013analysis,takaishi2018bias},} Heterogeneous Autoregressive (HAR)~\cite{corsi2009simple}, and Realized HAR \mbox{GARCH~\cite{huang2016modeling} models.}

In this study, we employ quantum circuit learning (QCL) introduced by Mitarai et al.~\mbox{\cite{mitarai2018quantum}},  
which has been proposed in the field of quantum computing, as an alternative approach to modeling volatility. Quantum computers aim to achieve computational speedup over classical methods by exploiting quantum-specific properties; in particular, their application is expected in the fields of quantum simulation, such as computational chemistry and materials science (see, e.g., \mbox{\cite{bauer2020quantum}}).
QCL employs parameterized quantum circuits (PQCs) for function approximation and classification tasks. 
Here, we attempt to approximate the volatility function utilizing the expressive power of QCL for function approximation, not computational speedup over classical methods using quantum computers.

Although QCL offers flexibility in designing quantum circuits, it does not require specifying a functional form for volatility dynamics.
Unlike traditional volatility models, the model learns a mapping from past information to future volatility directly through the optimization of quantum circuit parameters. Prior work~\cite{takaishi2025volatility} has applied QCL to synthetic time series generated by GARCH models, demonstrating that PQCs can reproduce some statistical properties of the underlying data.

Here, we extend this line of research by applying QCL to real financial data. Using an RV constructed from high-frequency Bitcoin price data, we train a single-qubit PQC to approximate the volatility time series. We then analyze the statistical properties of the time series generated by the trained quantum circuit, with a particular focus on multifractal characteristics. Specifically, we employ Multifractal Detrended Fluctuation Analysis (MFDFA)~\cite{kantelhardt2002multifractal} to estimate generalized Hurst exponents and investigate whether the QCL-generated time series reproduces empirical features such as anti-persistence and multifractality observed in realized volatility increments. Our results provide new insights into the expressive power of quantum circuits in modeling financial time series.

The remainder of this paper is organized as follows: Section \ref{sec2} reviews GARCH-type volatility models and discusses their interpretation as functional approximations to volatility dynamics. Section \ref{sec3} introduces the QCL framework and describes the single-qubit parameterized quantum circuit employed to approximate the volatility function. \mbox{Section \ref{sec4}} outlines the MFDFA methodology used to characterize the statistical properties of both empirical and model-generated time series. Section \ref{sec5} describes the data construction, including the computation of RV from high-frequency Bitcoin prices, and describes the parameter optimization procedure. Section \ref{sec6} presents the empirical results, focusing on the Hurst exponent, anti-persistence, and multifractal properties of the predicted return and volatility increment series. Finally, Section \ref{sec7} concludes with a discussion of the main findings, their implications, and directions for future research.

\section{GARCH Models}\label{sec2}
 
The
 GARCH model expresses the conditional volatility as a function of
past returns and past volatilities, which can be interpreted as a
truncated Taylor expansion of a general volatility function. Let
$\sigma_t^2$ denote the conditional volatility at time $t$ and $r_t$ the
corresponding return. Suppose the next period volatility $\sigma_{t+1}^2$
is determined by some function $f$ as
\begin{equation}
  \sigma_{t+1}^2 = f\!\left(\sigma_t^2,\, r_t\right), \label{eq:vol_general}
\end{equation}
and that the return is generated by
\begin{equation}
  r_t = \sigma_t \epsilon_t,
\end{equation}
where $\epsilon_t \sim \mathcal{N}(0,1)$ is an i.i.d.\ standard normal
random variable.
 
Because the true functional form of $f(\sigma_t^2, r_t)$ is unknown,
for practical purposes it must be approximated by some model. Expanding $f(\sigma_t^2, r_t)$
in a Taylor series around $\sigma_t^2 = 0$ and $r_t = 0$ gives
\begin{equation}
  f\!\left(\sigma_t^2, r_t\right)
  = f_{00}
  + f_{10}\,\sigma_t^2
  + f_{01}\,r_t
  + f_{11}\,\sigma_t^2 r_t
  + \tfrac{1}{2}f_{20}\!\left(\sigma_t^2\right)^2
  + \tfrac{1}{2}f_{02}\!\left(r_t\right)^2
  + \cdots, \label{eq:taylor}
\end{equation}
where the expansion coefficients are defined as
\begin{equation}
  f_{ij}
  = \left(\frac{\partial}{\partial\sigma_t^2}\right)^{\!i}
    \left(\frac{\partial}{\partial r_t}\right)^{\!j}
    f\!\left(\sigma_t^2, r_t\right)\bigg|_{\sigma_t^2,\,r_t=0}.
  \label{eq:taylor_coeff}
\end{equation}

If
 volatility is assumed to be symmetric with respect to the sign of
$r_t$, the coefficients $f_{01}$ and $f_{11}$ vanish. Retaining only
the lowest-order terms in $\sigma_t^2$ and $r_t$ yields
\begin{equation}
  f\!\left(\sigma_t^2, r_t\right)
  = f_{00} + f_{10}\,\sigma_t^2 + \tfrac{1}{2}f_{02}\!\left(r_t\right)^2,
  \label{eq:garch11_derivation}
\end{equation}
which corresponds to the well-known GARCH(1,1) model:
\begin{equation}
  \sigma_{t+1}^2 = \omega + \beta\,\sigma_t^2 + \alpha\,r_t^2.
  \label{eq:garch11}
\end{equation}
Here, $\alpha$, $\beta$, and $\omega$ correspond to Taylor expansion
coefficients. Since their true values are unknown, they are estimated by
fitting the model to historical time series data---equivalently, one is
estimating the expansion coefficients from the data.  
 
Volatility asymmetry with respect to the sign of $r_t$ is well
documented in equity price time series. Retaining the term proportional
to $r_t$ to account for this asymmetry gives
\begin{equation}
  f\!\left(\sigma_t^2, r_t\right)
  = f_{00} + f_{10}\,\sigma_t^2 + f_{01}\,r_t
    + \tfrac{1}{2}f_{02}\!\left(r_t\right)^2,
  \label{eq:qarch}
\end{equation}
which is the functional form of the QARCH
model~\cite{Sentana1995RES}. Increasing the order of the Taylor expansion
improves the approximation but also increases the number of free
parameters, making estimation increasingly difficult without necessarily
yielding commensurate gains in accuracy. An alternative is to
approximate $f$ by a class of functions other than polynomials.
Pad\'{e} approximants (In the finance literature, research applying Padé approximations to the approximation of probability distributions can be found in, for example, \cite{Nuyts2001Physica,chen2013empirical}.), 
for example, use rational functions and are known
to outperform Taylor series in many settings. A GARCH variant exploiting
this idea is the RGARCH model~\cite{takaishi2017rational}:
\begin{equation}
  f\!\left(\sigma_t^2, r_t\right)
  = \frac{\omega + \alpha\,r_t^2 + \beta\,\sigma_t^2}{1 + \gamma\,r_t},
  \label{eq:rgarch}
\end{equation}
and its stabilized version~\cite{takaishi2018volatility}:
\begin{equation}
  f\!\left(\sigma_t^2, r_t\right)
  = \frac{\omega + \alpha\,r_t^2 + \beta\,\sigma_t^2}{\exp(\gamma\,r_t)},
  \label{eq:rgarch2}
\end{equation}
where
 $\gamma$ captures volatility asymmetry. Setting $\gamma = 0$
reduces these to the GARCH(1, 1) model in Equation~\eqref{eq:garch11}.

Other functional forms have also been proposed~\cite{bera1993arch}. The Exponential GARCH
(EGARCH) model~\cite{Nelson1991Econ} operates on the logarithm of the variance,
while the APARCH model~\cite{ding1993long} treats
the power exponent of volatility as a free parameter. The diversity of
available models reflects the fact that each produces different
volatility estimates, and no single model dominates across all data sets
and asset classes.

\section{Quantum Circuit Learning}\label{sec3}

To model the volatility dynamics, this study employs the Quantum Circuit
Learning (QCL) approach proposed by Mitarai et al.~\mbox{\cite{mitarai2018quantum}}.
Within this framework, a parameterized quantum circuit is trained to
approximate the underlying volatility-generation process.

The objective is to construct a function that maps historical market
information to future volatility. Consistent with conventional GARCH-type
specifications, the explanatory variables consist of the lagged volatility
and lagged return. The volatility model is therefore represented as

\begin{equation}
v_t^2 = v(\boldsymbol{x}_{t-1},\boldsymbol{\theta}),
\end{equation}
where
$\boldsymbol{x}_{t-1}=(v_{t-1}^2,r_{t-1})$
denotes the input vector and
$\boldsymbol{\theta}$ represents the trainable circuit parameters.

The learning task consists of adjusting
$\boldsymbol{\theta}$ so that the model output reproduces the observed
volatility sequence. Let
$\sigma_i^2$
be the target volatility value and define the input \mbox{data as}
\begin{equation}
\boldsymbol{x}_i=(r_i,\sigma_i^2),
\end{equation}
where $r_i$ and $\sigma_i^2$ correspond to the return and volatility at
time $i$, respectively.

\begin{figure}[H]
\isPreprints{\centering}{} 
\begin{quantikz}[thin lines]
\lstick{$\ket{0}$}
& \gate{R_Y^r}
& \gate{R_Z^r}
& \gate{R_Y^v}
& \gate{R_Z^v}
& \gate{U(\boldsymbol{\theta})}
& \qw
& \meter{}
\end{quantikz}
\caption{Single-qubit
 quantum circuit used for volatility modeling.}
\label{fig:volatility_circuit}
\end{figure}
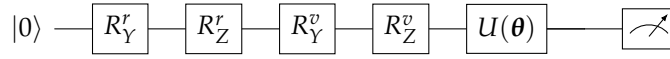

Figure~\mbox{\ref{fig:volatility_circuit}} presents the parameterized
single-qubit quantum circuit adopted in this study.
This model is a parsimonious specification with three parameters, and it is adopted based on comparisons with existing models. The GARCH(1,1) model, expressed in Equation~\mbox{\eqref{eq:garch11}}, is a simple three-parameter model; however, it is capable of capturing key characteristics of financial time series, such as volatility clustering, and is therefore widely used in empirical financial analysis. This suggests that it may be possible to describe essential features of the financial time series without introducing a large number of parameters. In this study, the model illustrated in Figure \ref{fig:volatility_circuit} is employed as the baseline specification.
It should be noted, however, that a three-parameter model has inherent limitations in terms of expressibility. Depending on the type of the financial time series under consideration, more expressive models may be required to adequately capture the underlying dynamics.

The trainable operation
$U(\boldsymbol{\theta})$
is expressed by the following two-dimensional unitary matrix:

\begin{equation}
U(\boldsymbol{\theta})
=
\begin{pmatrix}
\cos(\theta/2)
&
-e^{i\lambda}\sin(\theta/2)
\\[4pt]
e^{i\phi}\sin(\theta/2)
&
e^{i(\lambda+\phi)}\cos(\theta/2)
\end{pmatrix},
\end{equation}
where
$\boldsymbol{\theta}=(\theta,\lambda,\phi)$
contains the variational parameters to be optimized during training.

The classical input variables are encoded into the quantum state through
angle encoding using rotational gates $R_Y$ and $R_Z$. Following the
encoding scheme introduced in Ref.~\mbox{\cite{mitarai2018quantum}}, the gate
angles are defined as

\begin{align}
R_Y^r
&=
R_Y\!\left(\arcsin(r_i)\right),
\\
R_Z^r
&=
R_Z\!\left(\arccos(r_i^2)\right),
\\
R_Y^v
&=
R_Y\!\left(\arcsin(2\sigma_i^2-1)\right),
\\
R_Z^v
&=
R_Z\!\left(\arccos(\sigma_i^4)\right).
\end{align}

After application of the circuit, a measurement is performed in the
computational ($Z$) basis. The probability of observing the state
$\ket{0}$, denoted by $P_0$, is interpreted as the forecasted volatility
for the next period:

\begin{equation}
v_{i+1}^2=P_0.
\end{equation}

For numerical implementation, the probability $P_0$ is evaluated directly
from the quantum state vector using
Qiskit (IBM Qiskit:
 \url{https://www.ibm.com/quantum/qiskit}, accessed on 1 December 2025). Model training is carried out by optimizing the parameter vector
$\boldsymbol{\theta}$ so as to minimize the discrepancy between the
predicted and observed volatilities. The loss function is \mbox{defined as}

\begin{equation}
\mathcal{L}
= \frac1N
\sum_{i=1}^{N}
\left(
v_i^2-\sigma_i^2
\right)^2,
\end{equation}
where $N$ denotes the total number of observations used for estimation.


\section{Multifractal Analysis}\label{sec4}
To examine the multifractal characteristics of the analyzed time series,
the generalized Hurst exponent is estimated using the Multifractal
Detrended Fluctuation Analysis (MFDFA) framework. MFDFA, originally
proposed by Kantelhardt et al.~\mbox{\cite{kantelhardt2002multifractal}}, has
become a standard approach for detecting multifractal behavior in a wide
range of financial datasets (see, for example, \mbox{\cite{Jiang-Xie-Zhou-Sornette-2019-RPP}}). 
The main computational procedure is summarized below.

\begin{enumerate}
\item[(\romannumeral1)] Construction of the profile.

\end{enumerate}

Consider a time series
$\{x_t,\, t=1,\ldots,N\}$. The cumulative profile is obtained by
integrating the demeaned observations:

\begin{equation}
y(k)=\sum_{t=1}^{k}\left[x_t-\langle x\rangle\right],
\qquad k=1,\ldots,N,
\label{eq:profile}
\end{equation}
where $\langle x\rangle$ denotes the sample mean of the series.

\begin{enumerate}
\item[(\romannumeral2)] Estimation of local fluctuations.
\end{enumerate}

The profile $y(k)$ is partitioned into
$N_s=\lfloor N/s \rfloor$ consecutive non-overlapping segments, each of
length $s$. For every segment, a polynomial trend is fitted and removed,
after which the variance of the detrended profile is evaluated. For the
$\nu$-th segment ($\nu=1,\ldots,N_s$), the variance is given by

\begin{equation}
F^2(s,\nu)
=
\frac{1}{s}
\sum_{i=1}^{s}
\left[
y\bigl((\nu-1)s+i\bigr)-p_\nu(i)
\right]^2,
\label{eq:local_variance}
\end{equation}
where $p_\nu(i)$ denotes the fitted polynomial representing the local
trend. In the present analysis, a cubic polynomial is adopted,

\begin{equation}
p_\nu(i)=a+b\,i+c\,i^2+d\,i^3,
\end{equation}
with the coefficients estimated through least-squares fitting.

Since the series length $N$ is generally not an exact multiple of $s$,
part of the data may remain unused. To incorporate these observations,
the same segmentation procedure is repeated from the opposite end of the
series. For
$\nu=N_s+1, \ldots, 2N_s$, the detrended variance becomes

\begin{equation}
F^2(s,\nu)
=
\frac{1}{s}
\sum_{i=1}^{s}
\left[
y\bigl(N-(\nu-N_s-1)s-i+1\bigr)-p_\nu(i)
\right]^2.
\label{eq:local_variance_end}
\end{equation}

\begin{enumerate}
\item[(\romannumeral3)] Calculation of the fluctuation function.
\end{enumerate}

Using the collection of variances
$\{F^2(s,\nu)\}$, the fluctuation function of order $q$ is defined as

\begin{equation}
F_q(s)
=
\left\{
\frac{1}{2N_s}
\sum_{\nu=1}^{2N_s}
\left[F^2(s,\nu)\right]^{q/2}
\right\}^{1/q}.
\label{eq:fluctuation}
\end{equation}

For a process characterized by long-range scale-invariant
correlations, the fluctuation function follows the power-law relation

\begin{equation}
F_q(s)\propto s^{h(q)},
\label{eq:scaling}
\end{equation}
where $h(q)$ denotes the generalized Hurst exponent. Its value is
estimated from the slope of the linear relationship in the logarithmic
representation of Equation~\mbox{\eqref{eq:scaling}}. When $q=2$, MFDFA reduces to
the conventional Detrended Fluctuation Analysis
(DFA)~\mbox{\cite{peng1994mosaic}}, and the exponent $h(2)$ corresponds to the
standard Hurst exponent~\mbox{\cite{hurst1951long}}.

The expression in Equation~\mbox{\eqref{eq:fluctuation}} is not valid for $q=0$.
Therefore, the fluctuation function for this case is evaluated as

\begin{equation}
F_0(s)
=
\exp\left\{
\frac{1}{4N_s}
\sum_{\nu=1}^{2N_s}
\ln\left[F^2(s,\nu)\right]
\right\}.
\label{eq:fluctuation_q0}
\end{equation}

\begin{enumerate}
\item[(\romannumeral4)] Estimation of the generalized Hurst exponent.
\end{enumerate}

The generalized Hurst exponent $h(q)$ is obtained from the scaling
behavior described in Equation~\mbox{\eqref{eq:scaling}}. Throughout this study,
the calculation is carried out over the interval
$q\in[-5,5]$.
Large values of $|q|$ may lead to numerical instability in the estimation of moments used in the fluctuation function. Jiang et al.~\mbox{\cite{Jiang-Xie-Zhou-Sornette-2019-RPP}} pointed out that, for low-frequency financial data such as daily observations, it is preferable to restrict the analysis to $|q| \leq 6$. Following their recommendation, the multifractal measures were computed over the range
$q = [-5,5]$ in this study.

A time series is classified as monofractal when $h(q)$ remains
independent of $q$, whereas variation in $h(q)$ across different values
of $q$ indicates multifractality. For an ideal Gaussian random
process, $h(q)=1/2$ for all $q$. Consequently, deviations from this
behavior provide evidence of multifractal structure and have often been
associated with departures from market efficiency.

Several additional quantities can be derived from the generalized Hurst
exponent~\cite{kantelhardt2002multifractal}. One of them is the
multifractal scaling exponent,

\begin{equation}
\tau(q)=q\,h(q)-1,
\label{eq:tau}
\end{equation}
which becomes a linear function of $q$ in the monofractal case. The
singularity spectrum $f(\alpha)$ is obtained through a Legendre
transformation of $\tau(q)$:

\begin{align}
\alpha(q)
&=
\frac{d\tau(q)}{dq},
\label{eq:alpha}
\\
f\!\left(\alpha(q)\right)
&=
q\,\alpha(q)-\tau(q).
\label{eq:falpha}
\end{align}

Substituting Equation~\eqref{eq:tau} into the above expressions yields

\begin{align}
\alpha(q)
&=
h(q)+q\frac{dh(q)}{dq},
\label{eq:alpha_h}
\\
f\!\left(\alpha(q)\right)
&=
q\bigl[\alpha(q)-h(q)\bigr]+1.
\label{eq:falpha_h}
\end{align}

 
\section{Data and Parameter Optimization}\label{sec5}
 
In this study, daily realized volatility $\mathrm{RV}_t$ is
constructed from Bitcoin prices traded on the Bitstamp exchange, using
price data sampled at five-minute intervals. While higher sampling
frequencies improve the precision of RV estimates, they
also amplify the influence of microstructure noise~\cite{zhou1996high,bandi2006separating,hansen2006realized}. The five-minute
sampling interval is therefore adopted as a standard compromise between
estimation accuracy and noise contamination, consistent with the
recommendation in the existing literature~\cite{liu2015does}. The five-minute log-return is
defined as
\begin{equation}
  r_{t,i} = \ln P_{t,i+1} - \ln P_{t,i}, \label{eq:return}
\end{equation}
where $P_{t,i}$ denotes the five-minute price on day $t$ for $i=1,...,1440/5$. The daily realized volatility $\mathrm{RV}_t$ on day $t$
is then given by
\begin{equation}
  \mathrm{RV}_t = \sum_{i=1}^{n} r_{t,i}^2,
  \label{eq:RV}
\end{equation}
\textls[-25]{where $n$ is the number of intraday observations; for five-minute
sampling, $n = 1440/5 = 288$.}
 
The data cover the period {from 31 August 2020 to 1 March 2023
 (UTC)
(913 days, or approximately two and a half years)} obtained from the Bitstamp exchange (Bitstamp exchange: \url{https://www.bitstamp.net/}, accessed on 3 March 2023). 

\begin{figure}[H]
\vspace{1.5cm}
\isPreprints{\centering}{} 
\includegraphics[width=9 cm]{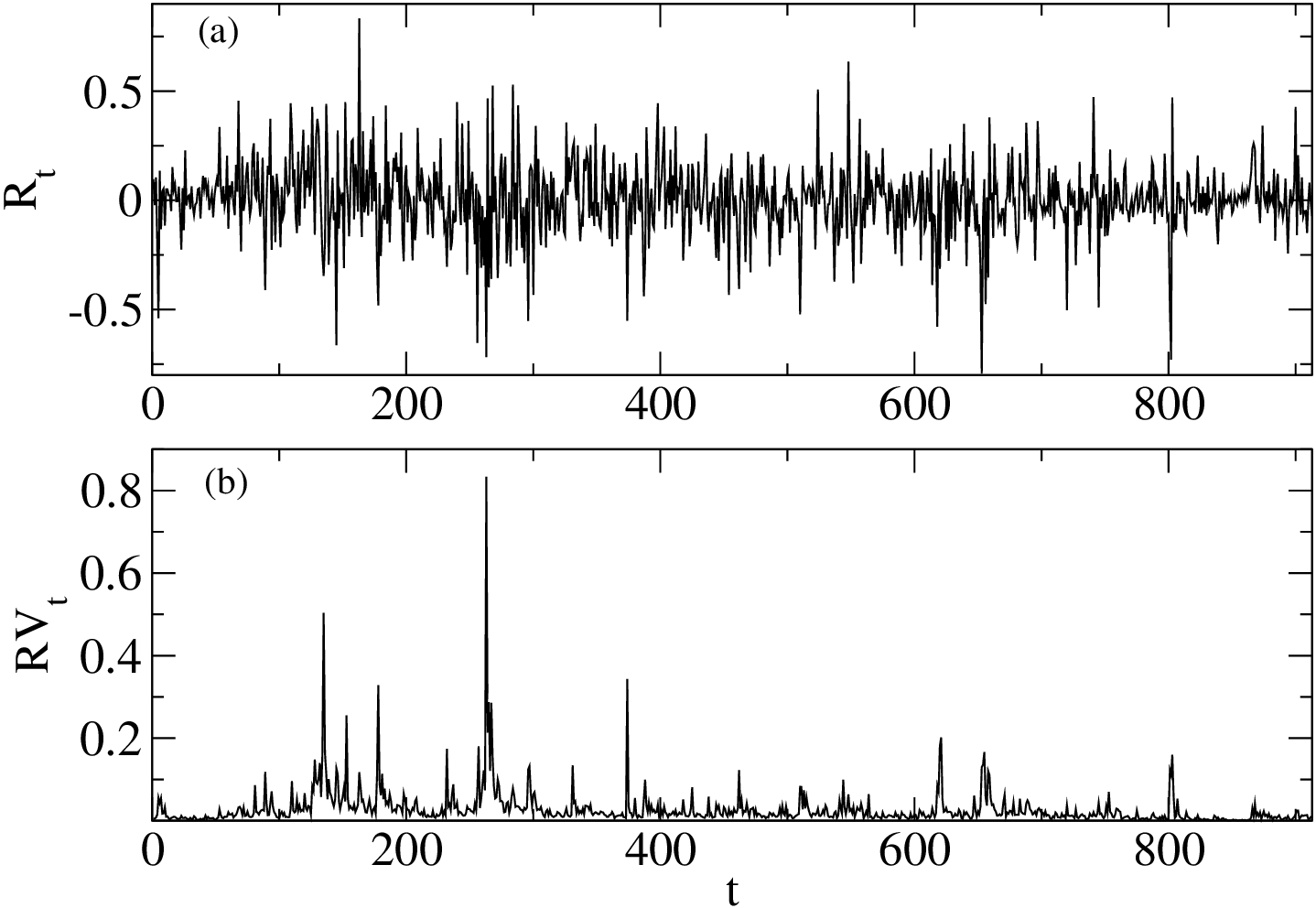}
\caption{(\textbf{a}) Return
 time series $R_t$. (\textbf{b}) Realized volatility time series $RV_t$.\label{fig2}}
\end{figure}

We construct a 5-min price time series from the obtained high-frequency data and compute the realized volatility according to Equation (\mbox{\ref{eq:RV}}). If no transaction data are available within a 5-min interval, we use the price from the previous interval.
Figure \ref{fig2} plots (a) the daily return series
$R_t$, computed from daily prices $P_t$ via Equation~\eqref{eq:return}, and
\mbox{(b) the} daily realized volatility $\mathrm{RV}_t$.

In this study, the model parameters are estimated using the entire realized volatility dataset to investigate its statistical and generative properties. Accordingly, we do not explicitly divide the data into training, validation, and testing subsets. The primary objective of this work is not to evaluate predictive performance, but rather to examine the model's ability to reproduce the empirical characteristics of the realized volatility.

The return series $R_t$ and the RV series
$\mathrm{RV}_t$ are used as inputs to the quantum circuit. Because
input values are encoded as rotation angles, both series are
pre-scaled: $R_t$ is mapped to $[-1/c,\, 1/c]$ and $\mathrm{RV}_t$ to
$[0,\, 1/c]$, with $c = 1.2$. 
{In the formulation of Equation~(11), the input data are given by 
$\boldsymbol{x_{t-1}}=(R_{t-1},RV_{t-1})$. The vector $\boldsymbol{x_{t-1}}$ is fed into the QCL model, 
and the predicted value for the next time step, $\overline{\mathrm{RV}}_t$, 
is obtained as $\overline{\mathrm{RV}}_t = P_0$ according to Equation~(17). 
The resulting $\overline{\mathrm{RV}}_t$ is then used to evaluate the loss function.}
The parameters of $U(\boldsymbol{\theta})$
are optimized by minimizing the loss function
\begin{equation}
  \mathcal{L}
  = \frac{1}{T}\sum_{t=1}^{T}
    \Bigl(\mathrm{RV}_t - \overline{\mathrm{RV}}_t\Bigr)^2,
  \label{eq:cost}
\end{equation}
where $\overline{\mathrm{RV}}_t$ is the quantum circuit output at time
$t$ and $T$ is the total number of observations. Parameters are optimized
using the COBYLA algorithm.

\begin{figure}[H]
\vspace{1.5cm}
\isPreprints{\centering}{} 
\includegraphics[width=10.0 cm]{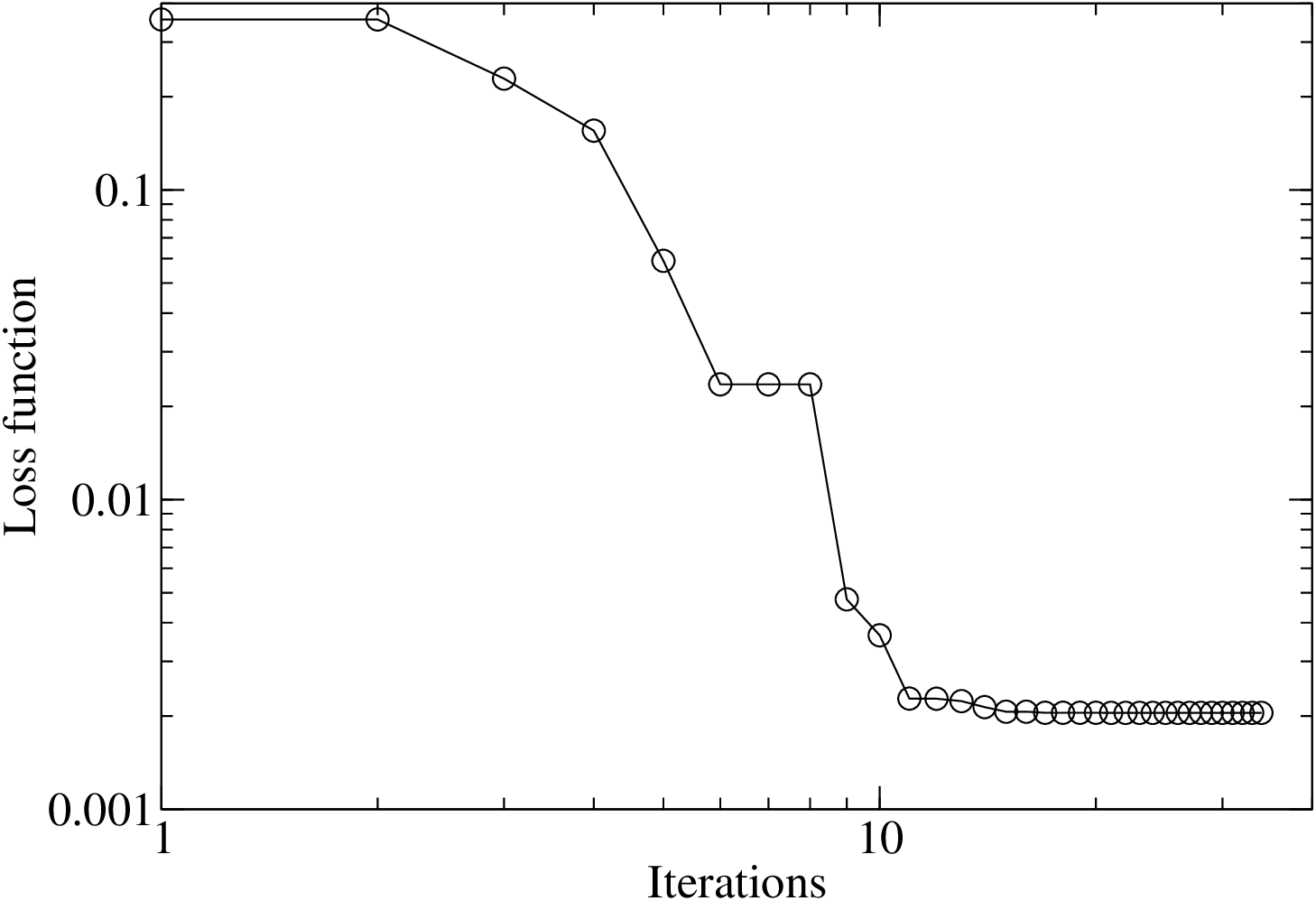}
\caption{Loss function during optimization by the COBYLA algorithm as a function of iteration. }
\label{fig:loss_convergence}
\end{figure}

We initialized the parameters with random values and executed the COBYLA algorithm using several different random initializations; while the algorithm generally exhibited good convergence, we observed poor convergence in certain cases depending on the initial values. 
{When convergence was achieved, the loss function converged to similar values.}
Figure \ref{fig:loss_convergence} illustrates the relationship between the loss function and the number of iterations for a converged case.

\begin{figure}[H]
\vspace{1.5cm}
\isPreprints{\centering}{} 
\includegraphics[width=10.0 cm]{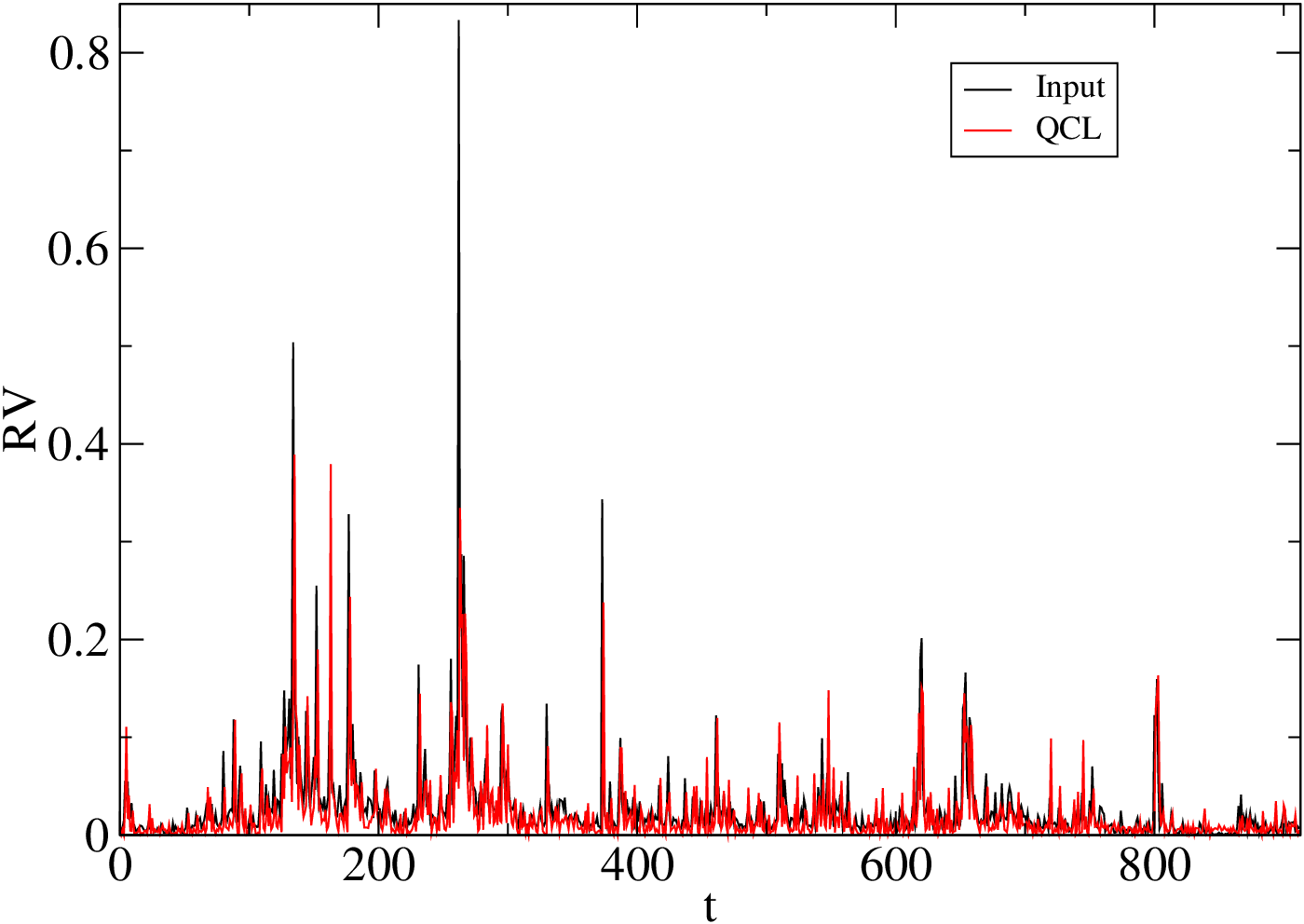}
\caption{(Black line) Empirical time series data of $RV_t$ used as input for the QCL. (Red line) QCL outputs $\overline{RV_t}$ after optimization.}\label{fig4}
\end{figure}

Figure \ref{fig4} 
 compares the input series $\mathrm{RV}_t$ with the circuit
output $\overline{\mathrm{RV}}_t$ after optimization. The circuit output
reproduces the broad temporal variation in the input $\mathrm{RV}_t$.

\section{Properties of the Predicted Time Series}\label{sec6}
 
To investigate the statistical properties of the QCL model,
we generated 50,000 time steps of predicted data
by the QCL model with the optimized parameters $\boldsymbol{\theta}$.
The QCL model first predicts the next realized volatility $\overline{\mathrm{RV}}_t$ using the input data from the previous period. Based on the predicted realized volatility, the corresponding return $\overline{R}_t$ is then computed as follows:
\begin{equation}
  \overline{R}_t = \overline{\mathrm{RV}}_t^{1/2}\,\epsilon_t,
  \label{eq:return_sim}
\end{equation}
where $\epsilon_t \sim \mathcal{N}(0,1)$.

Using the predicted return $\overline{R}_t$ and realized volatility $\overline{\mathrm{RV}}_t$ as new input data, the model proceeds to forecast the subsequent return and realized volatility.

The formulation in Equation (\ref{eq:return_sim}) reflects the empirical observation that returns normalized by realized volatility following a Gaussian distribution~\cite{andersen2000exchange,andersen2007no,takaishi2012finite}. Consequently, the fluctuations in realized volatility allow the return distribution to exhibit fat tails. For instance, if the realized volatility follows an inverse-gamma distribution (The empirical results support that the distribution of realized volatility follows an inverse gamma distribution~\mbox{\cite{takaishi2010analysis}}.), the resulting return distribution corresponds to a Student's t-distribution.

\begin{figure}[H]
\vspace{1.5cm}
\isPreprints{\centering}{} 
\includegraphics[width=9.0 cm]{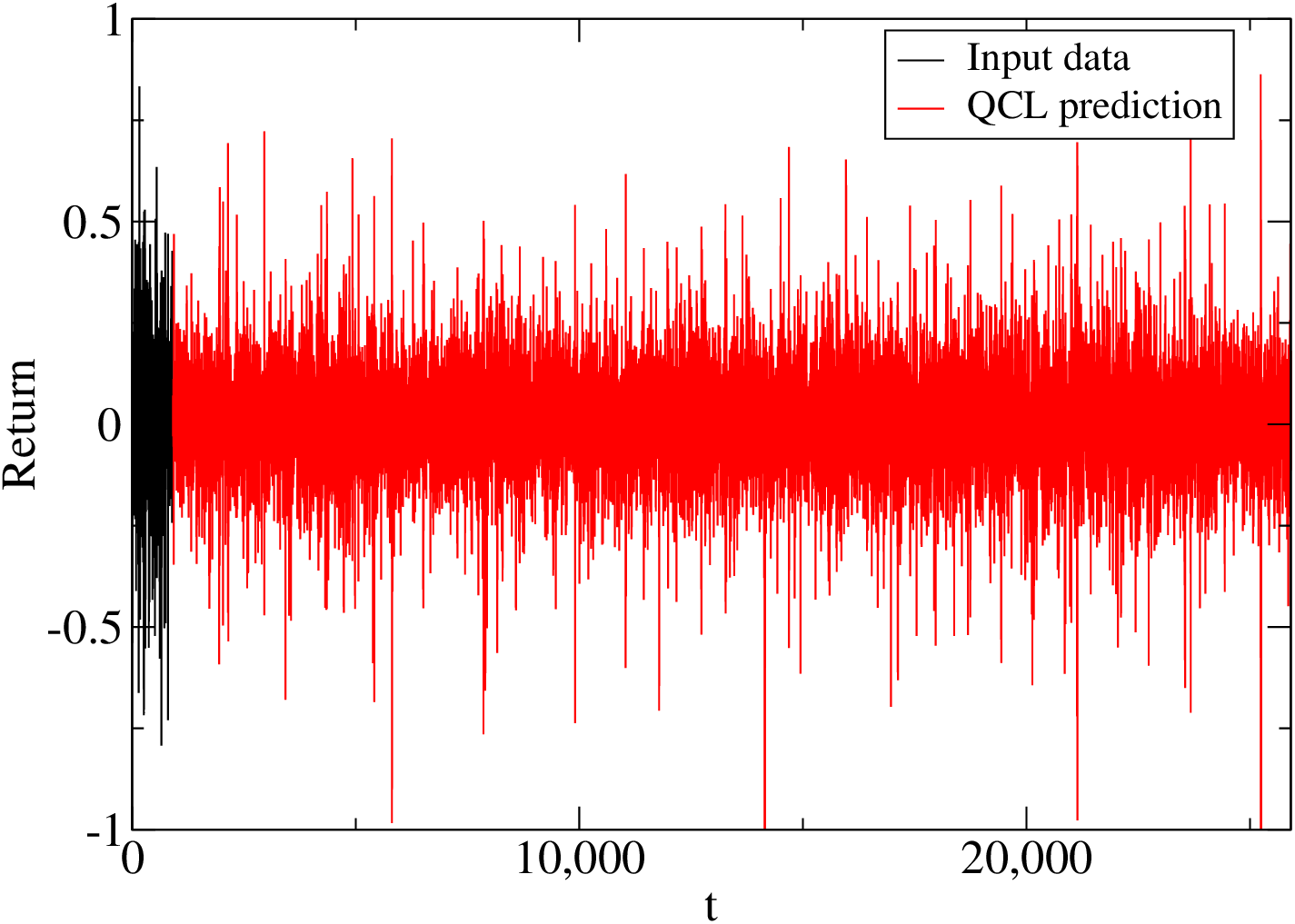}
\caption{The
 predicted return
series $\overline{R_t}$ (red line) and the input data $R_t$ (black line).
\label{fig5}}
\end{figure}

\begin{figure}[H]
\vspace{1.5cm}
\isPreprints{\centering}{} 
\includegraphics[width=9.0 cm]{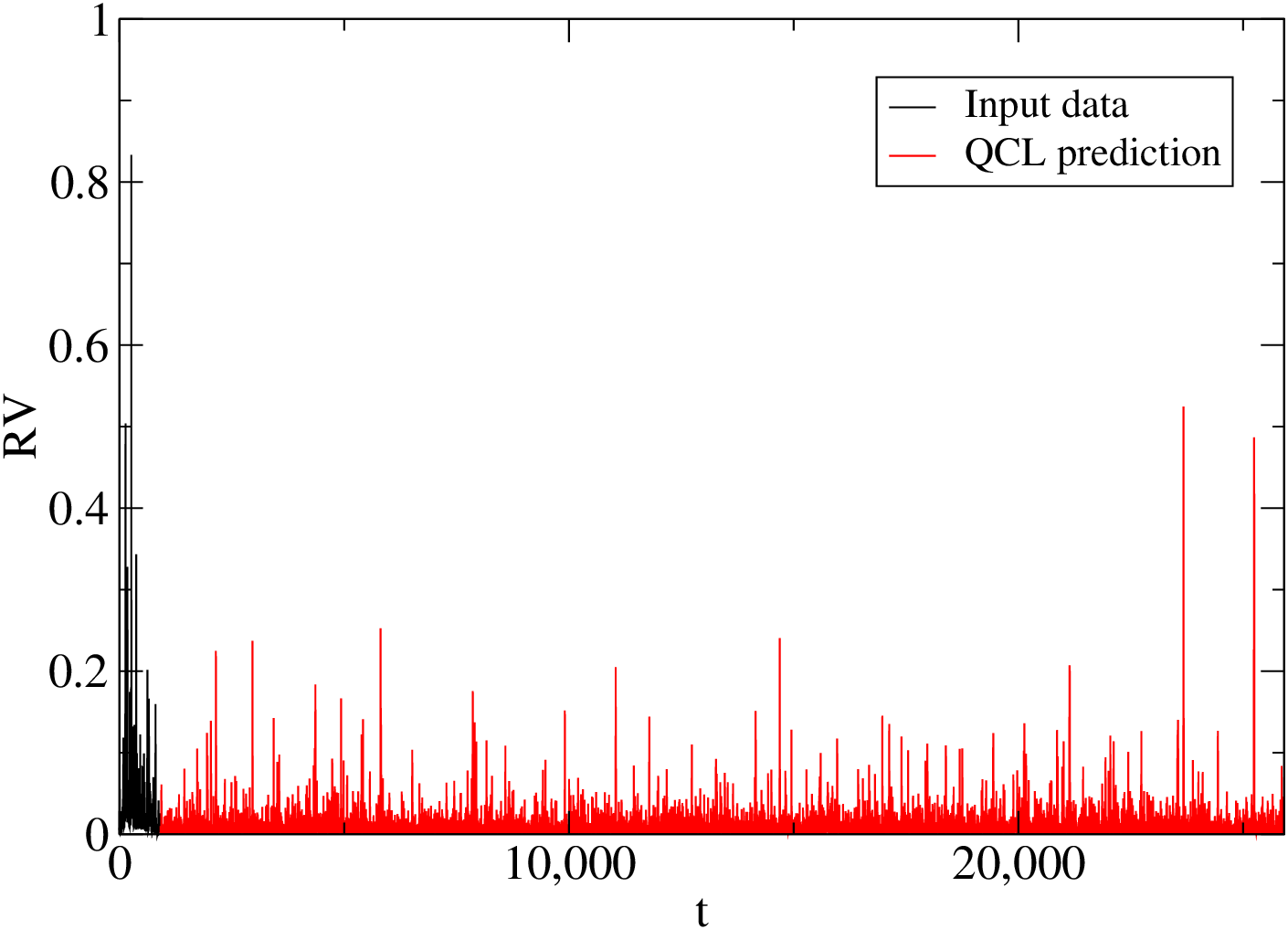}
\caption{The 
 predicted volatility series $\overline{\mathrm{RV}}_t$ (red line) and
the input $\mathrm{RV}_t$ (black line).
\label{fig6}}
\end{figure}

 
Figure \ref{fig5} shows the first 25,000 observations of the predicted return
series $\overline{R_t}$ (red line), together with the 913 observations of
the input data $R_t$ (black line) used for parameter optimization. 
Similarly, Figure \ref{fig6} displays the
predicted volatility series $\overline{\mathrm{RV}}_t$ (red line) and
the input $\mathrm{RV}_t$ (black line).

\begin{figure}[H]
\vspace{1.5cm}
\isPreprints{\centering}{} 
\includegraphics[width=9.0 cm]{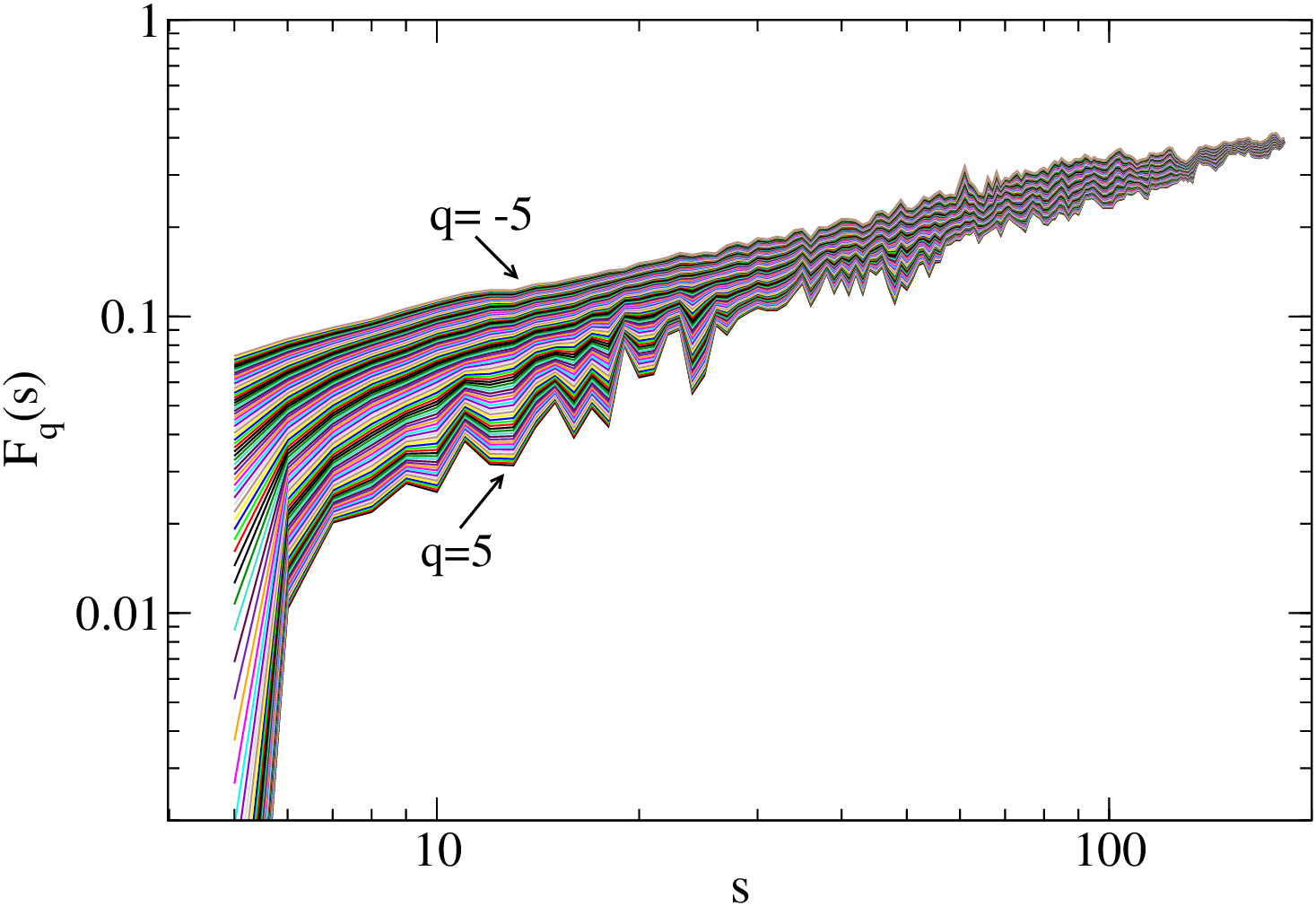}
\caption{The
 fluctuation function obtained from the empirical return time series. 
The fluctuation functions are plotted with a step size of $\Delta q=0.1$ in $q=[-5,5]$.}
\label{fig:fluctuation}
\end{figure}

The generalized Hurst exponent $h(q)$ is obtained as the scaling exponent of the fluctuation function{,} as defined in Equation (\mbox{\ref{eq:scaling}}). Figure~\mbox{\ref{fig:fluctuation}} shows a log--log plot of the fluctuation function for the Bitcoin return time series as a representative example. In this study, we computed the fluctuation function for $q$ in the range $[-5, 5]$ with a step size of $\Delta q = 0.02$. In Figure~\mbox{\ref{fig:fluctuation}}, for the sake of clarity, we only plot the results with a step size of $\Delta q = 0.1$, demonstrating linearity in the region of large $s$. We performed the linear fitting in the region $s \geq 20$ to estimate the scaling exponent $h(q)$.

The generalized Hurst exponent $h(q)$ of the 50,000-observation time series is
estimated using a rolling window approach with a window size of 913
observations and a step size of 365 observations. 
To compare with random time series, we generated 20 surrogate time series by randomly shuffling the data within each window. We then computed the generalized Hurst exponent for each surrogate time series and averaged the results.

\begin{figure}[H]
\vspace{1.5cm}
\isPreprints{\centering}{} 
\includegraphics[width=9.0 cm]{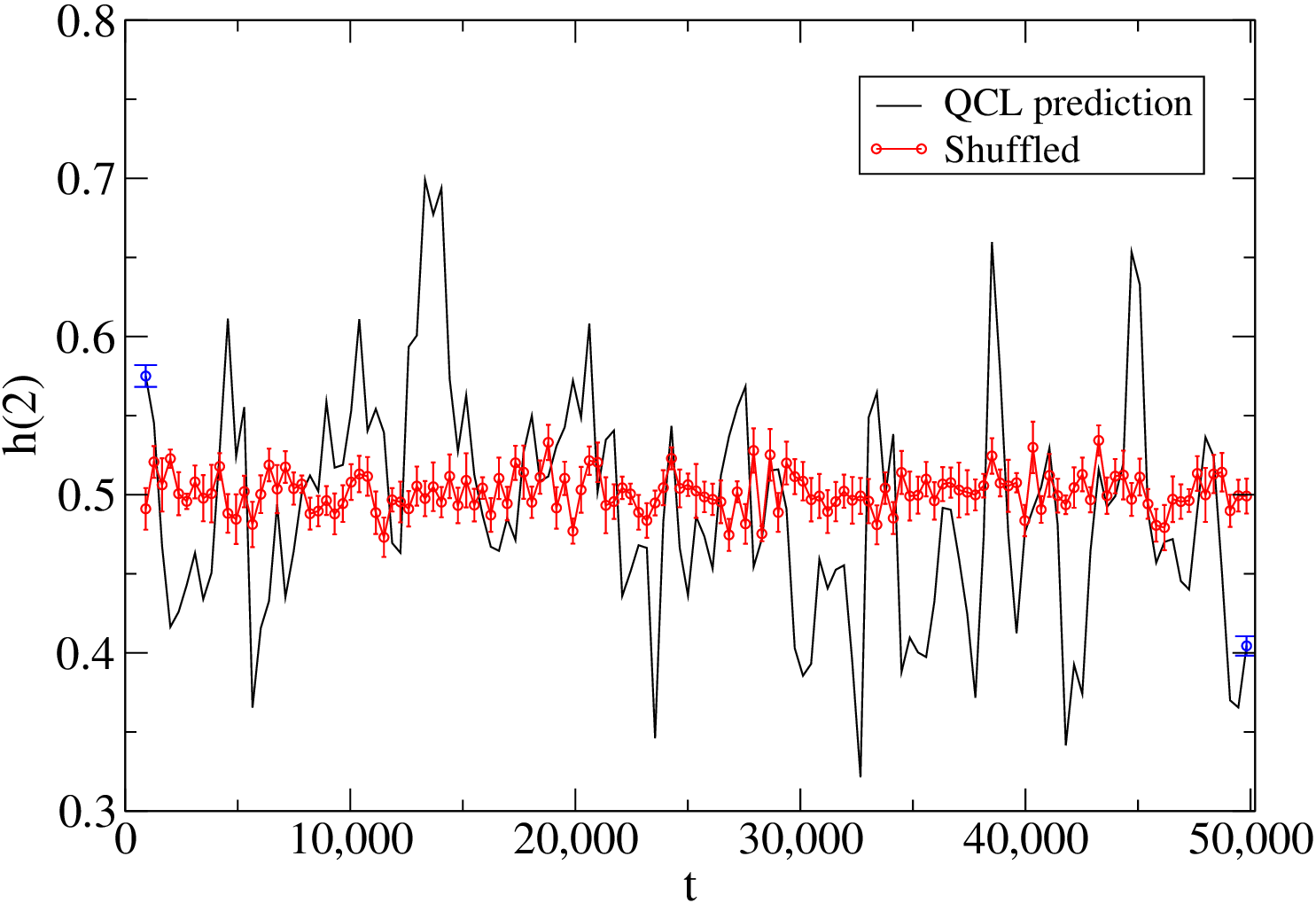}
\caption{Time
 variation in the Hurst exponent $h(2)$ for the
predicted and shuffled return series. The blue symbols with error bars in the figure indicate the typical magnitude of the fitting errors.}
\label{fig8}
\end{figure}

Figure \ref{fig8} shows the time variation in the Hurst exponent $h(2)$ for the
predicted and shuffled return series, analyzed by the rolling window method. 
The values fluctuate around $0.5$, indicating that the
predicted return series is approximately random (Empirical studies have shown that, in most developed markets, the Hurst exponent of asset returns is close to 0.5, corresponding to a random time series; however, it has also been reported that some developed markets exhibit anti-persistent behavior, with Hurst exponents below 0.5~\cite{MATTEO2005827}.), consistent with the
empirical behavior of Bitcoin returns reported in recent results~\cite{takaishi2019market}. The
$h(2)$ value of the empirical return series used as input data is
also close to $0.5$ (Table \ref{tab1}).
{It should be noted that, since the return in the QCL model is defined by Equation~(34), it is expected that the Hurst exponent takes a value close to 0.5; therefore, the present result simply confirms the generation of return time-series data based on Equation~(34).}

\begin{table}[H] 
\caption{$h(2)$ for empirical (input) return, IRV, and IAR time series.}\label{tab1}

\begin{tabularx}{\textwidth}{CC}
\toprule
 &   \boldmath{$h(2)$} \\
\midrule
Return	&  0.504(5)		 \\
Return (Shuffled) &  0.498(11)			 \\
\midrule
IRV		&  0.155(2) 		 \\
IRV (Shuffled) &  0.516(11)			\\
\midrule
IAR		&  0.030(1) 		 \\
IAR (Shuffled) &  0.494(14)		 \\
\bottomrule
\end{tabularx}
\end{table}

\begin{figure}[H]
\vspace{1.5cm}
\isPreprints{\centering}{} 
\includegraphics[width=9.0 cm]{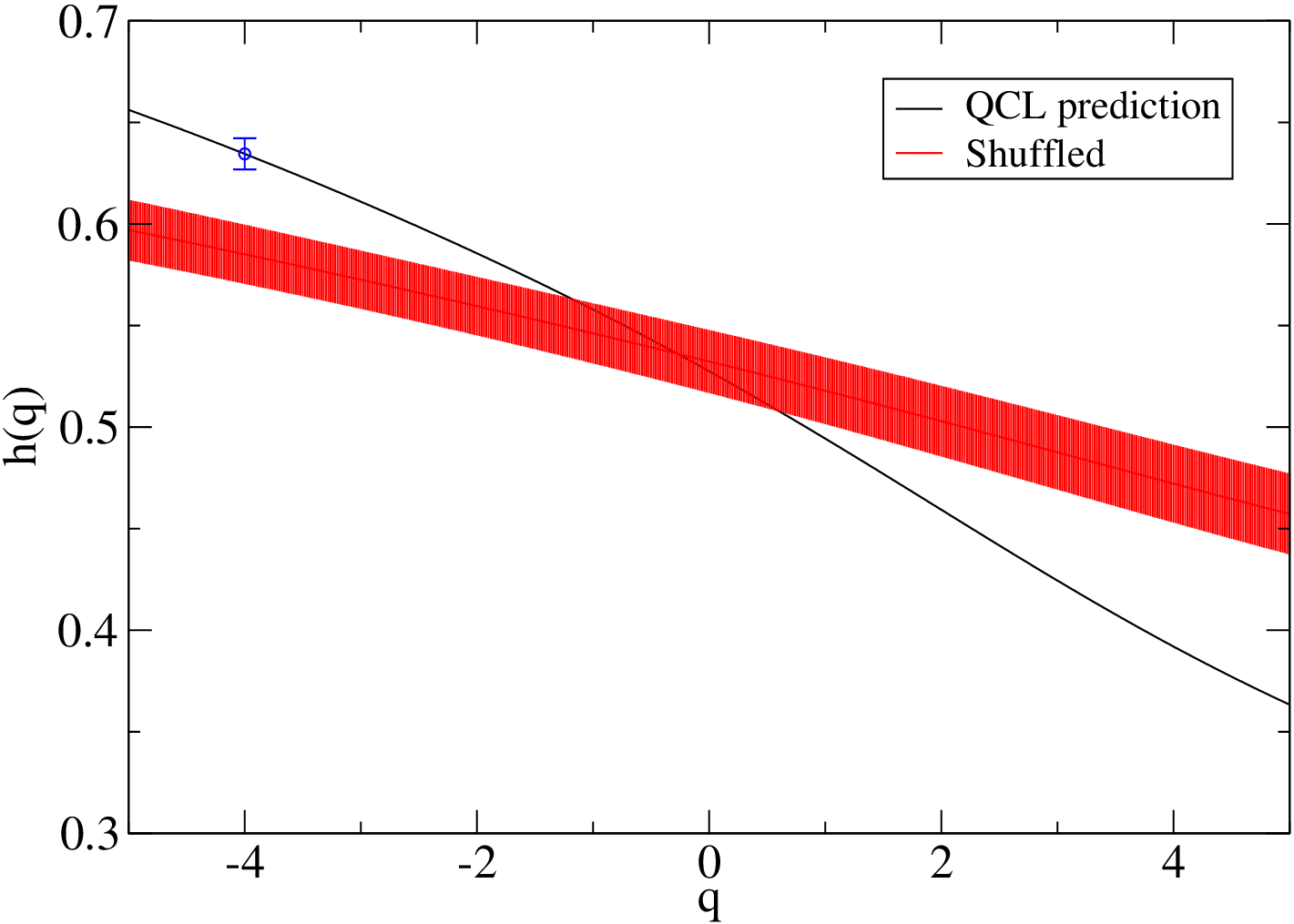}
\caption{$h(q)$ of
 the QCL-predicted and shuffled return time series in a representative window.
The blue symbol with error bars in the figure indicates the typical magnitude of the fitting error.}
\label{fig9}
\end{figure}

\begin{figure}[H]
\vspace{1.5cm}
\isPreprints{\centering}{} 
\includegraphics[width=9.0 cm]{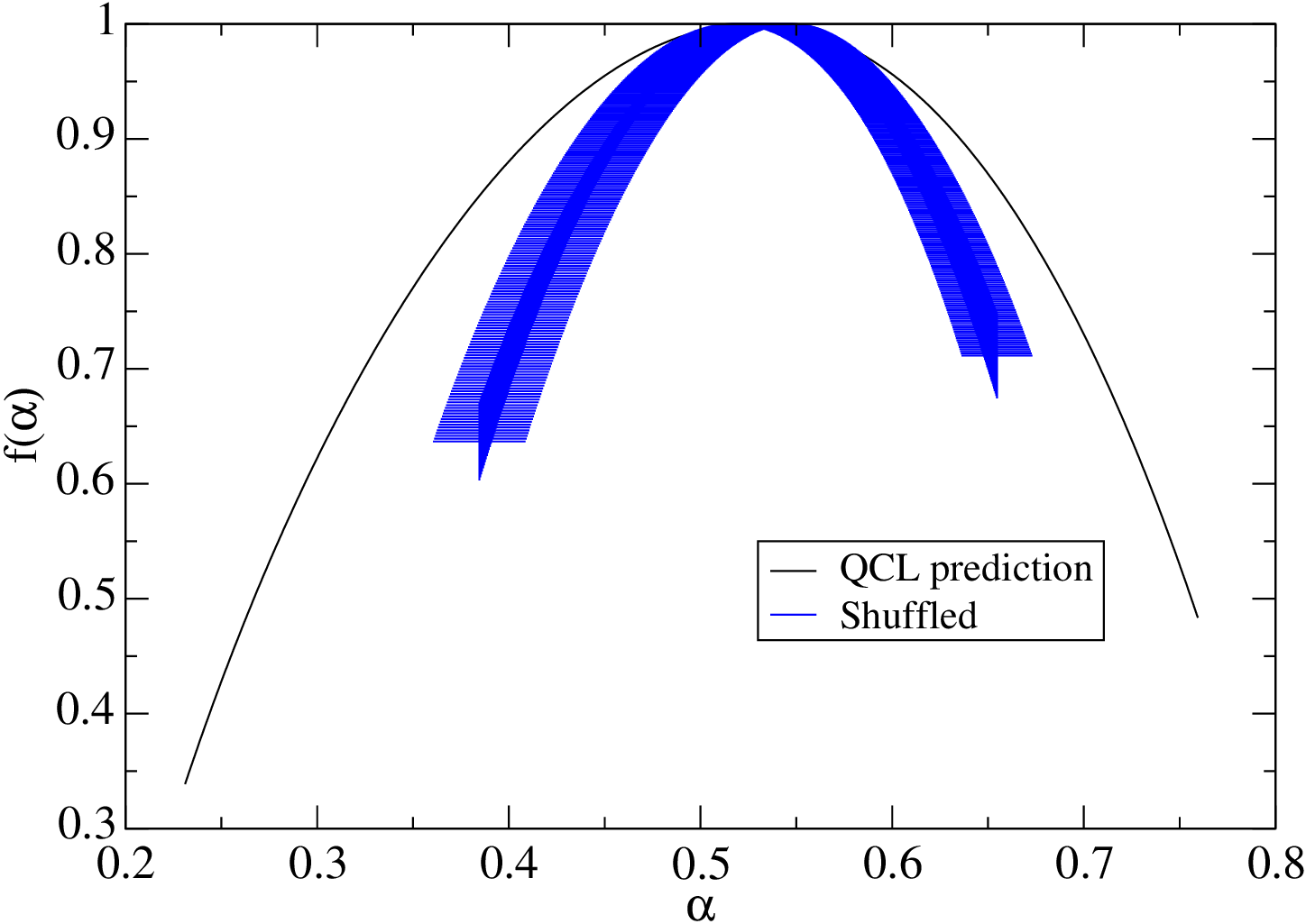}
\caption{$f(\alpha)$ of 
 the QCL-predicted and shuffled return time series in a representative window.
\label{fig10}}
\end{figure}

\begin{figure}[H]
\vspace{1.5cm}
\isPreprints{\centering}{} 
\includegraphics[width=9.0 cm]{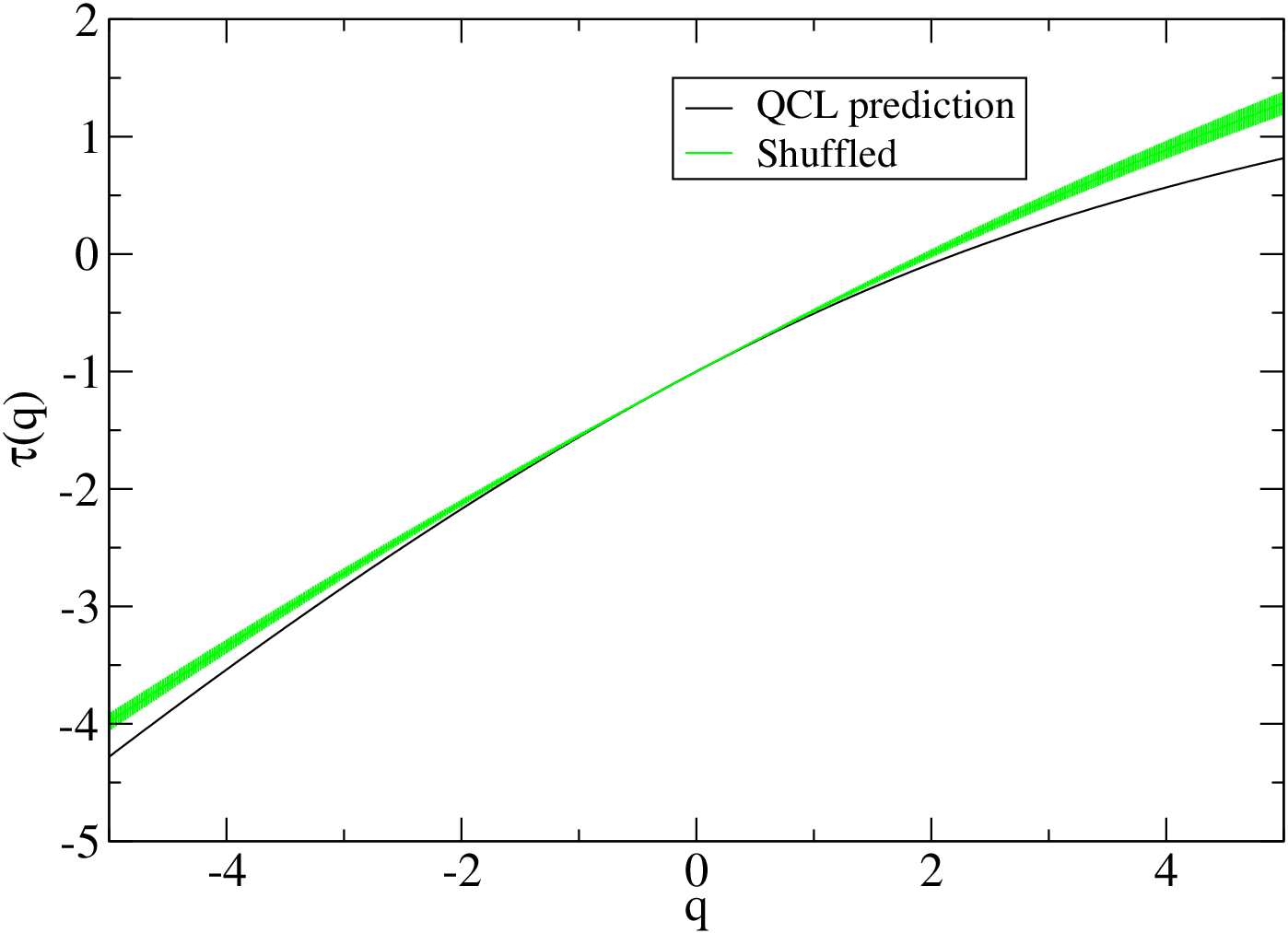}
\caption{$\tau(q)$ of
 the predicted return time series in a representative window.
\label{fig11}}
\end{figure}

Figures \ref{fig9}--\ref{fig11} display $h(q)$, $\alpha(q)$, and $\tau(q)$ for the QCL-predicted return
series in a representative window.
The variation in $h(q)$ with $q$ seems to show 
multifractality in the predicted return series, and  after shuffling, the
$q$-dependence of $h(q)$ is reduced but not eliminated. Both $f(\alpha)$ and $\tau(q)$ also exhibit similar multifractal properties.

\begin{figure}[H]
\vspace{1.5cm}
\isPreprints{\centering}{} 
\includegraphics[width=9.0 cm]{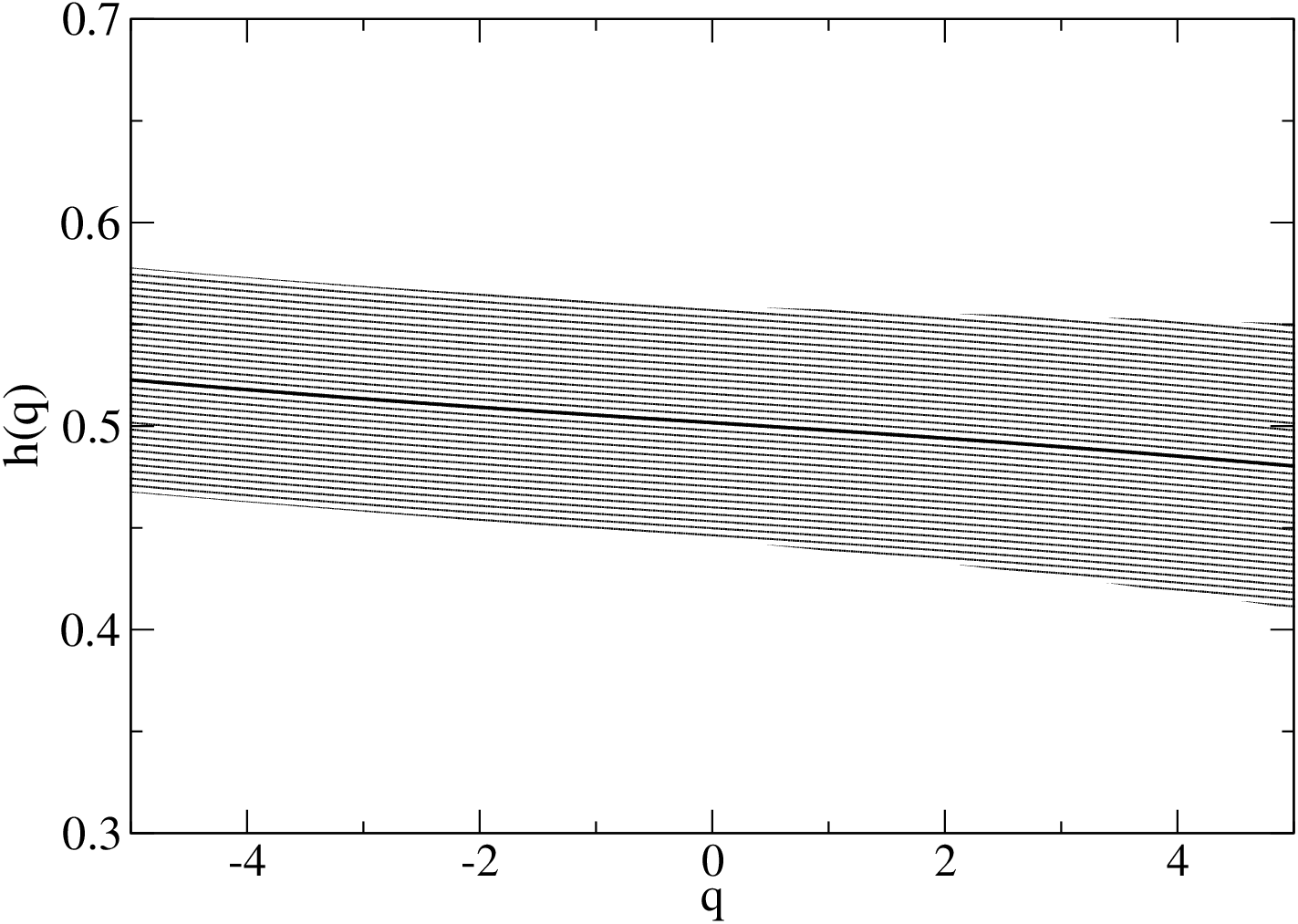}
\caption{$h(q)$ of
 Gaussian random time series from 200 samples. The width in the figure represents the standard deviation.}
\label{fig12}
\end{figure}

\begin{figure}[H]
\vspace{1.5cm}
\isPreprints{\centering}{} 
\includegraphics[width=9.0 cm]{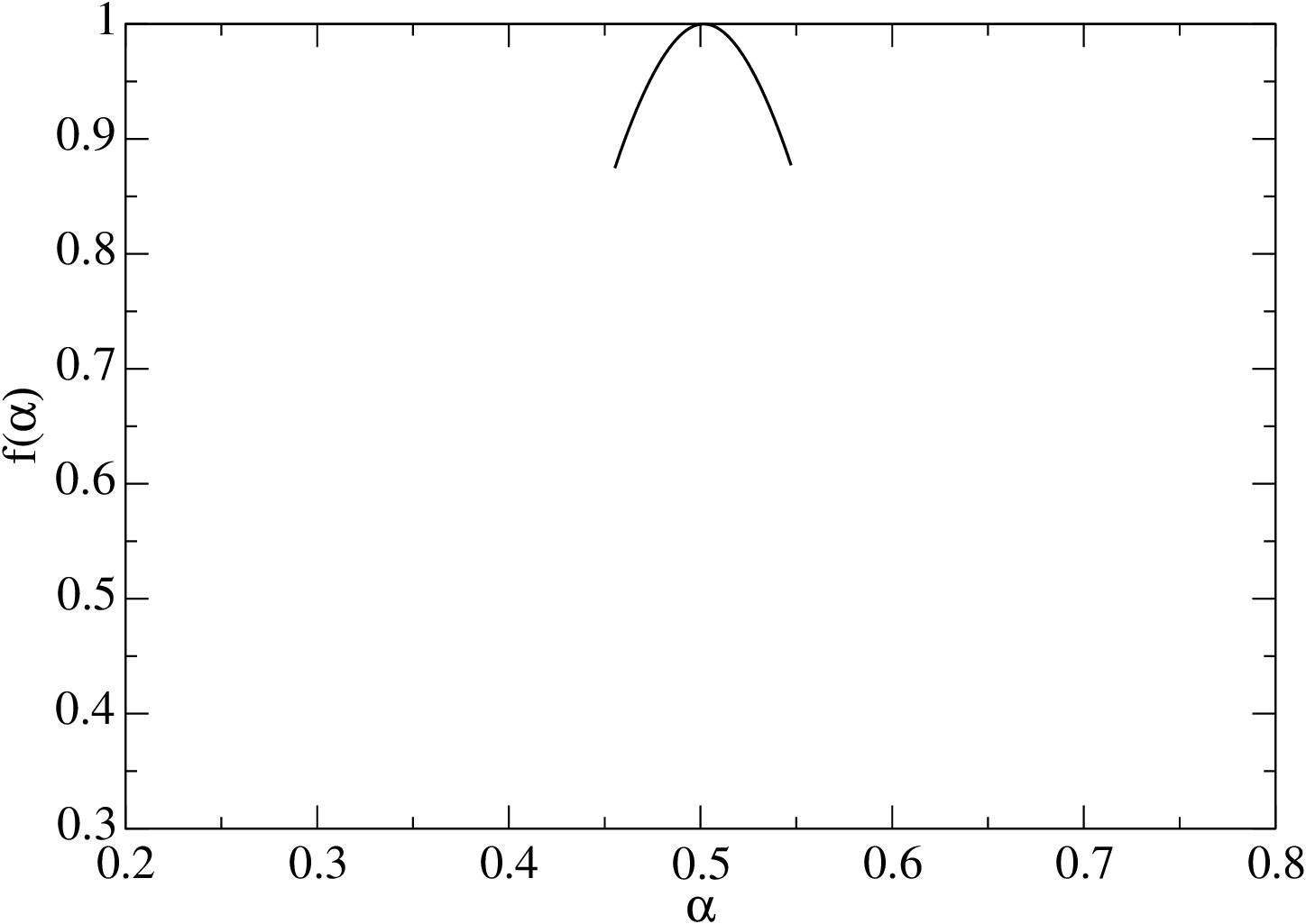}
\caption{$f(\alpha)$ of 
  Gaussian random time series.  $f(\alpha)$ was obtained from $h(q)$ shown in Figure \ref{fig12} numerically.}
\label{fig13}
\end{figure}

\begin{figure}[H]
\vspace{1.5cm}
\isPreprints{\centering}{} 
\includegraphics[width=9.0 cm]{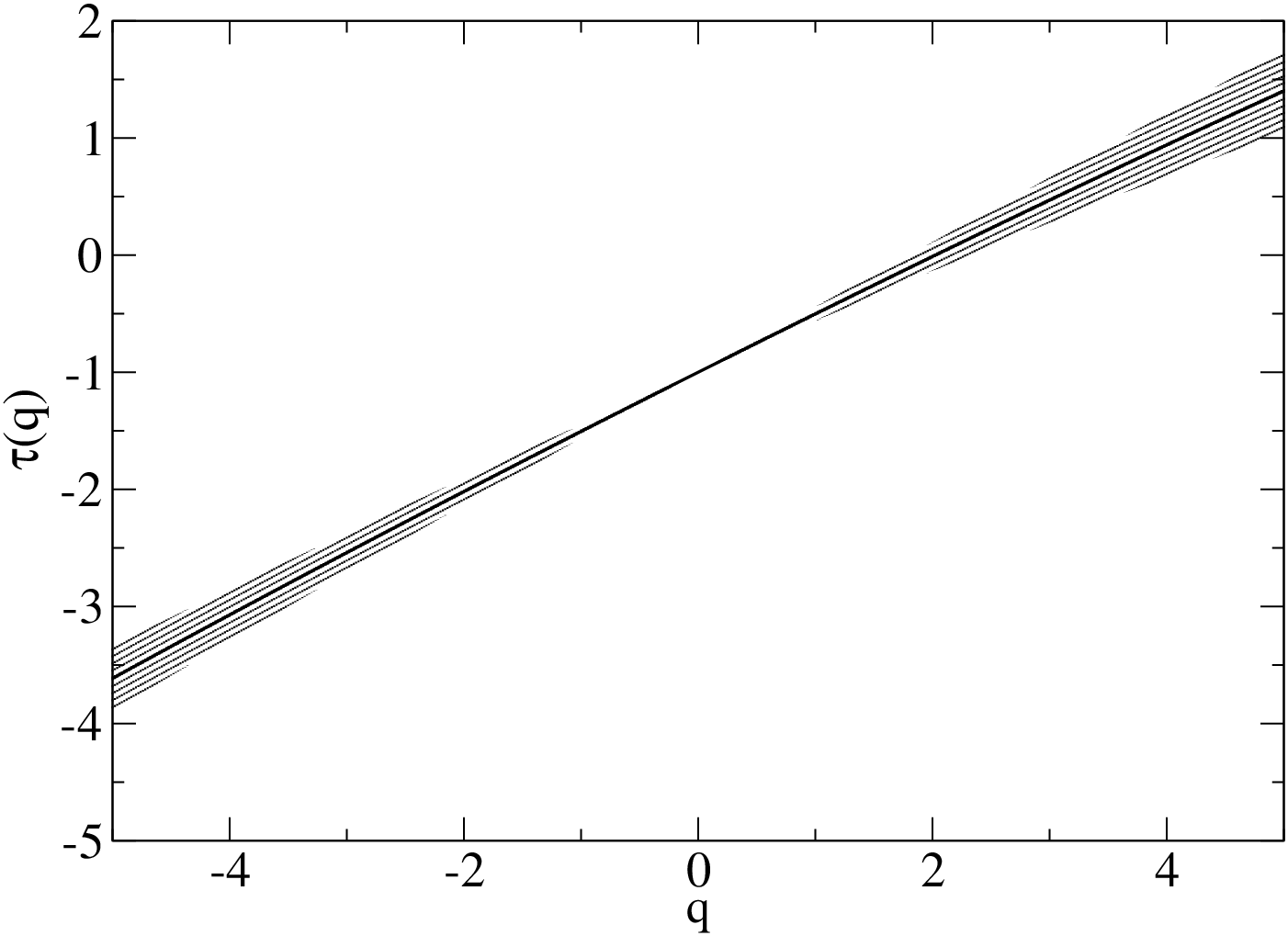}
\caption{$\tau(q)$ of
 Gaussian random time series from 200 samples. The width in the figure represents the standard deviation.}
\label{fig14}
\end{figure}

However, it is known that finite-size effects in the time series of finite length can lead to the emergence of spurious multifractality~\mbox{\cite{zhou2012finite,grech2012multifractal,grech2013multifractal}}. To estimate the magnitude of these effects for the time series length used in this study, we generated 200 Gaussian time series with a length of $n = 913$, computed their $h(q)$, $\alpha(q)$, and $\tau(q)$ values, and plotted the results in Figures \ref{fig12}--\ref{fig14}. For comparison, the scales in these figures were set to be identical to those in Figures \ref{fig9}--\ref{fig11}. It is observed that while Gaussian time series are inherently monofractal, they exhibit multifractality due to finite-size effects. Comparing Figures \ref{fig9}--\ref{fig11} with \mbox{Figures \ref{fig12}--\ref{fig14},} the degree of multifractality in the original time series is smaller in the Gaussian case, suggesting that the results presented in {Figures \ref{fig9}--\ref{fig11}} indicate the possible presence of multifractality beyond what can be attributed to finite-size effects.

It has been reported that the fat-tailed nature of the distribution in a time series can act as a source of multifractality~\mbox{\cite{kantelhardt2002multifractal}} (Table \ref{tab2} shows the kurtosis of return, IRV and IAR time series analyzed in this study.). Although shuffling a time series eliminates temporal correlations and is expected to allow for the separation of contributions from temporal correlations and distribution shapes as the origin of multifractality, recent studies have reported that true multifractality arises solely from non-linear temporal correlations~\mbox{\cite{kwapien2023genuine,kluszczynski2025disentangling}}. Consequently, it is difficult to discuss the origin of multifractality based solely on the shuffling process. Therefore, this study only reports the results for the original and shuffled time series.


{From Table \ref{tab2}, we observe that the kurtosis of both the empirical return data and the QCL model predictions exceeds 3, the value for a Gaussian distribution, indicating the presence of fat-tailed behavior. However, the kurtosis of the QCL model is larger than that of the empirical data. This discrepancy suggests that the single-qubit QCL model employed in this study may not fully capture certain properties of the empirical data and may overestimate the fat-tailed behavior.}

\begin{table}[H]
\caption{Kurtosis of return, IRV, and IAR time series.\label{tab2}}
\begin{tabularx}{\textwidth}{CCC}
			\toprule
		& \textbf{Input (Empirical) Data}	& \textbf{QCL Prediction}    \\
			\midrule
Return	& 5.79(60) 		& 15.7(33)			\\
IRV   & 4.43(87)	   & 4.55(7)			\\
IAR   & 4.40(42)	   & 4.35(10)			\\
                                
			\bottomrule
		\end{tabularx}
\end{table}

\begin{figure}[H]
\vspace{1.5cm}
\isPreprints{\centering}{} 
\includegraphics[width=9.0 cm]{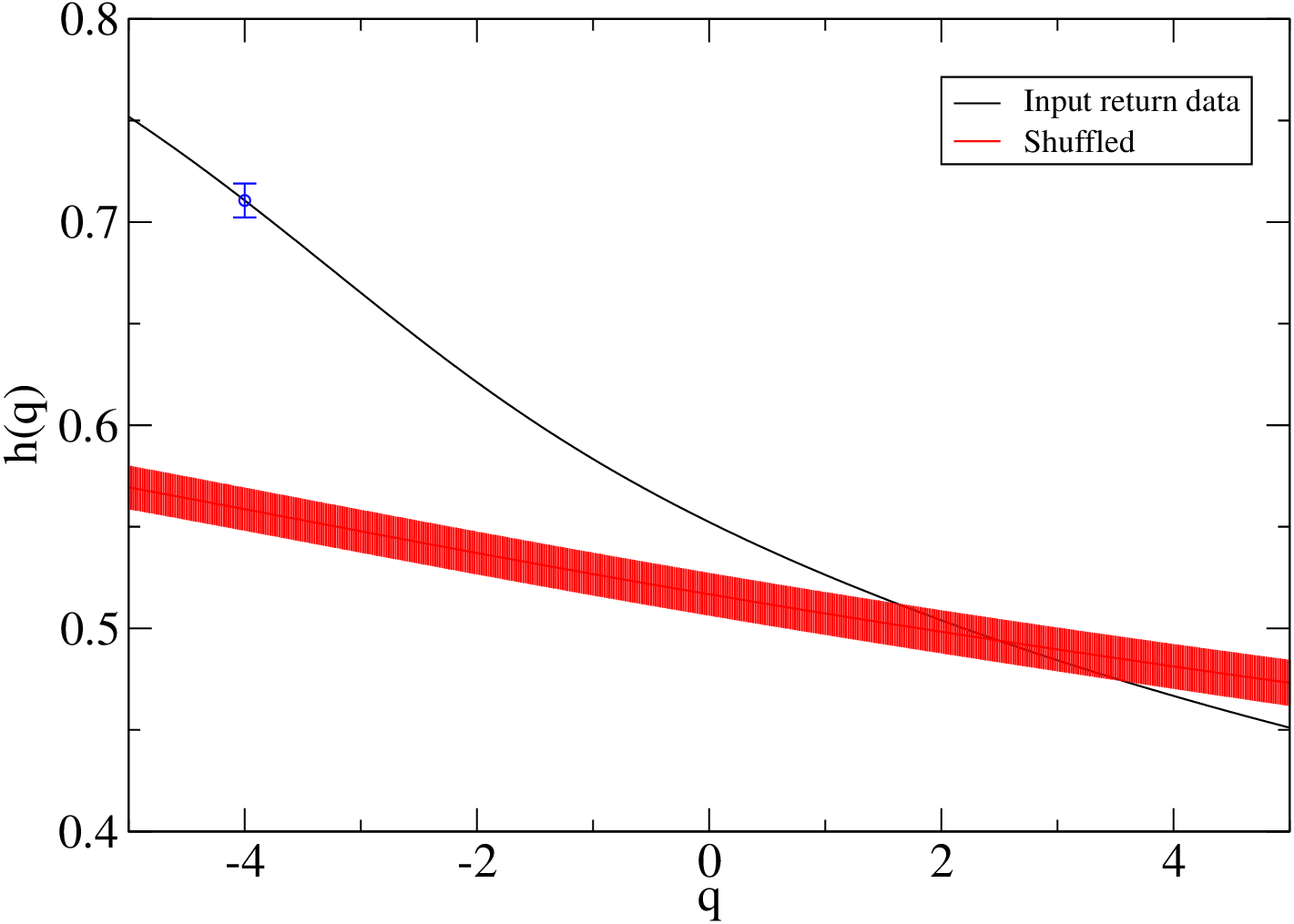}
\caption{$h(q)$ of 
 the empirical (input) return time series.
The blue symbol with error bars in the figure indicates the typical magnitude of the fitting error.}
\label{fig15}
\end{figure}

\begin{figure}[H]
\vspace{1.5cm}
\isPreprints{\centering}{} 
\includegraphics[width=9.0 cm]{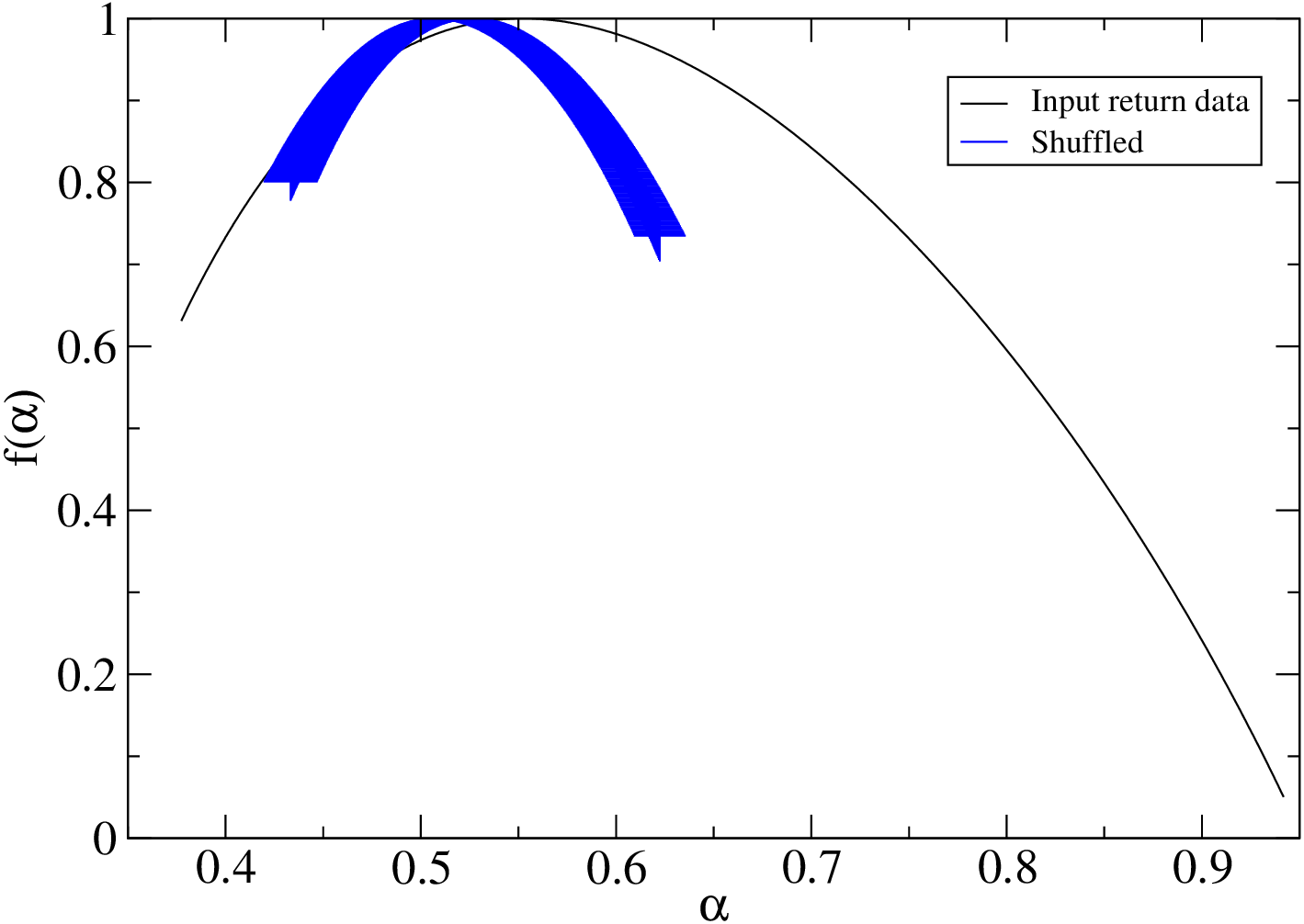}
\caption{$f(\alpha)$ of the empirical (input) return time series.
\label{fig16}}
\end{figure}

\begin{figure}[H]
\vspace{1.5cm}
\isPreprints{\centering}{} 
\includegraphics[width=9.0 cm]{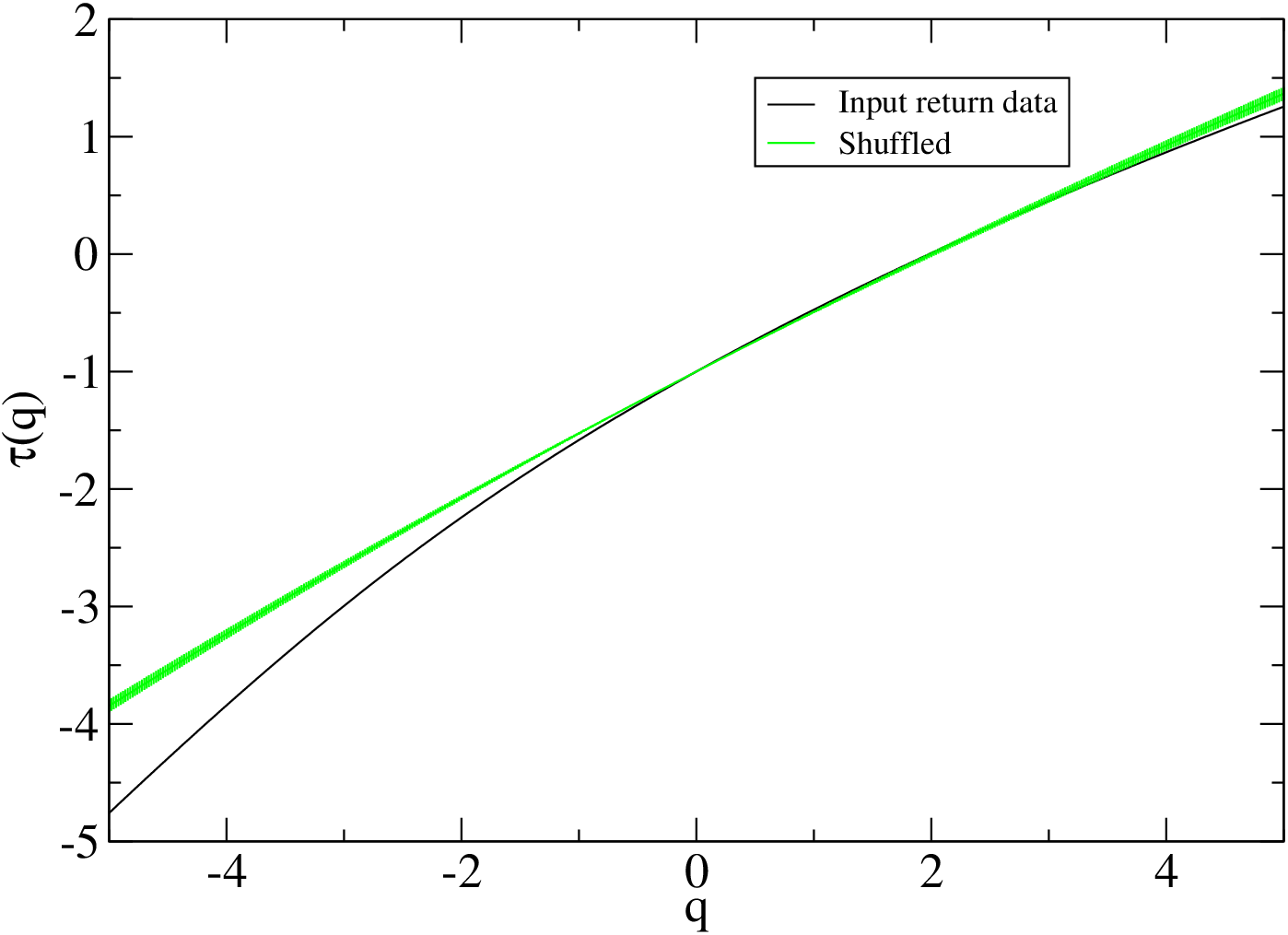}
\caption{$\tau(q)$
 of the empirical (input) return time series.
\label{fig17}}
\end{figure}

Figures \ref{fig15}--\ref{fig17} display  $h(q)$, $f(\alpha)$, and  $\tau(q)$ for the empirical return time series used as input. The multifractal characteristics of the empirical data closely resemble those of the time series generated by the QCL model, demonstrating that the model  reproduces the qualitatively similar multifractal features of empirical returns.

\vspace{-9pt}

Next, the same multifractal analysis is applied to the increment series of realized
volatility (IRV), defined as
\begin{equation}
  \mathrm{IRV}_t = \ln\mathrm{RV}_t - \ln\mathrm{RV}_{t-1}.
  \label{eq:irv}
\end{equation}
  
\begin{figure}[H]
\vspace{1.5cm}
\isPreprints{\centering}{} 
\includegraphics[width=9.0 cm]{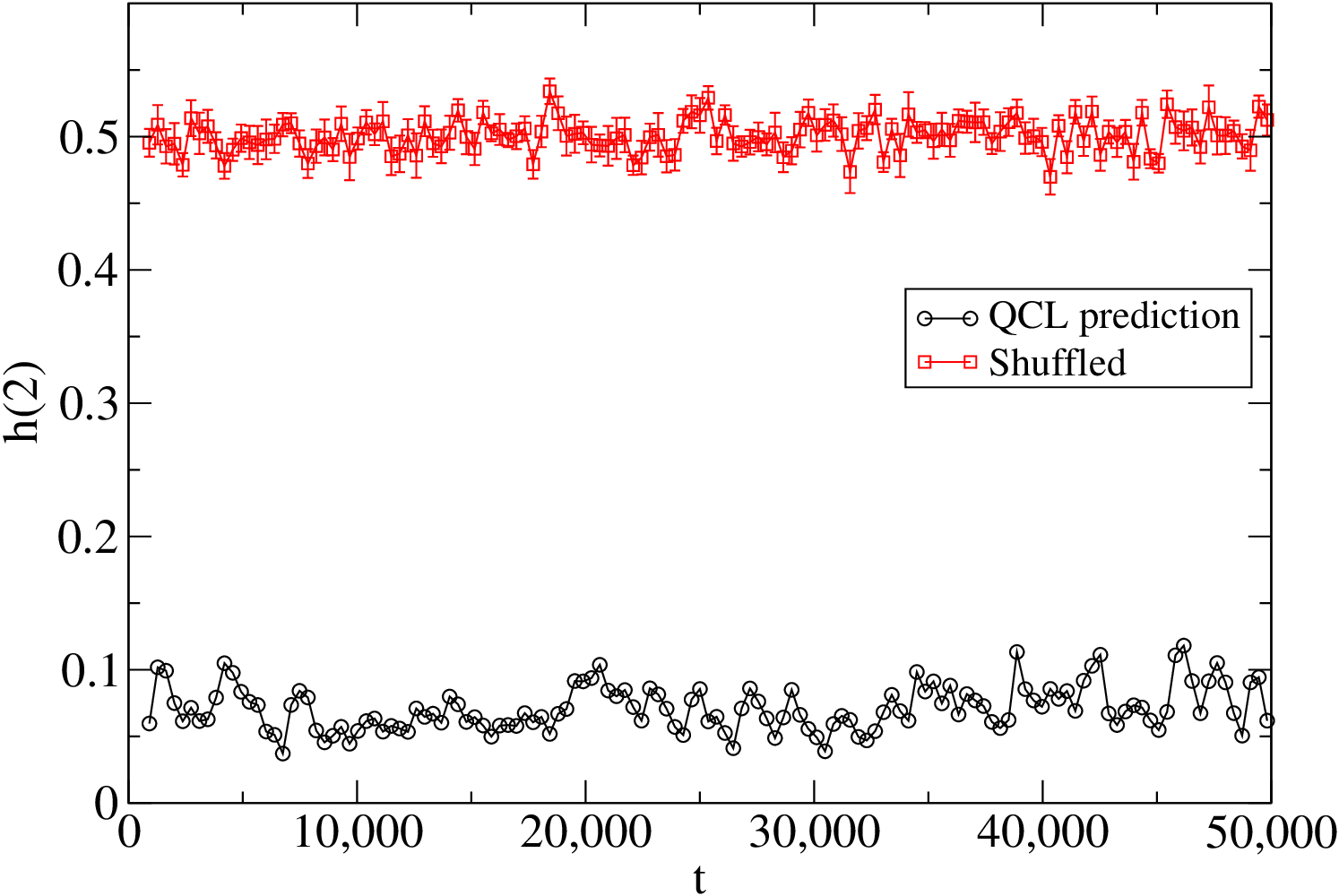}
\caption{Hurst 
 exponent $h(2)$ of the IRV time series and those of the shuffled time series.
For QCL predictions, the typical magnitude of the fitting error is on the order of the symbol size or smaller.}
\label{fig18}
\end{figure}

 Figure \ref{fig18} shows the Hurst exponent $h(2)$ of the IRV series, with
values in the range $h(2) \approx 0.05$--$0.1$ (on average $\sim 0.07$), well below the random
benchmark of $0.5$. This clearly confirms the presence of anti-persistence, which is consistent with the rough-volatility characteristics observed in the real financial data of Bitcoin~\cite{takaishi2020rough}.
After shuffling, $h(2)$ reverts to approximately $0.5$,
demonstrating that the anti-persistence is attributable to the temporal
correlation structure of the time series.


 \vspace{-9pt}

RV requires high-frequency data for accurate construction; however, when such data are unavailable, the absolute value of daily returns is often used as a proxy for volatility. In relation to RV, the square of the absolute daily return corresponds to the case in which RV is constructed using only a single intraday observation. Let $|R_t|$ denote the absolute daily return. The increment series of absolute returns, denoted $\mathrm{IAR}_t$, is defined as
\begin{equation}
    \mathrm{IAR}_t = \ln |R_t| - \ln |R_{t-1}|.
\end{equation}

\begin{figure}[H]
\vspace{1.5cm}
\isPreprints{\centering}{} 
\includegraphics[width=9.0 cm]{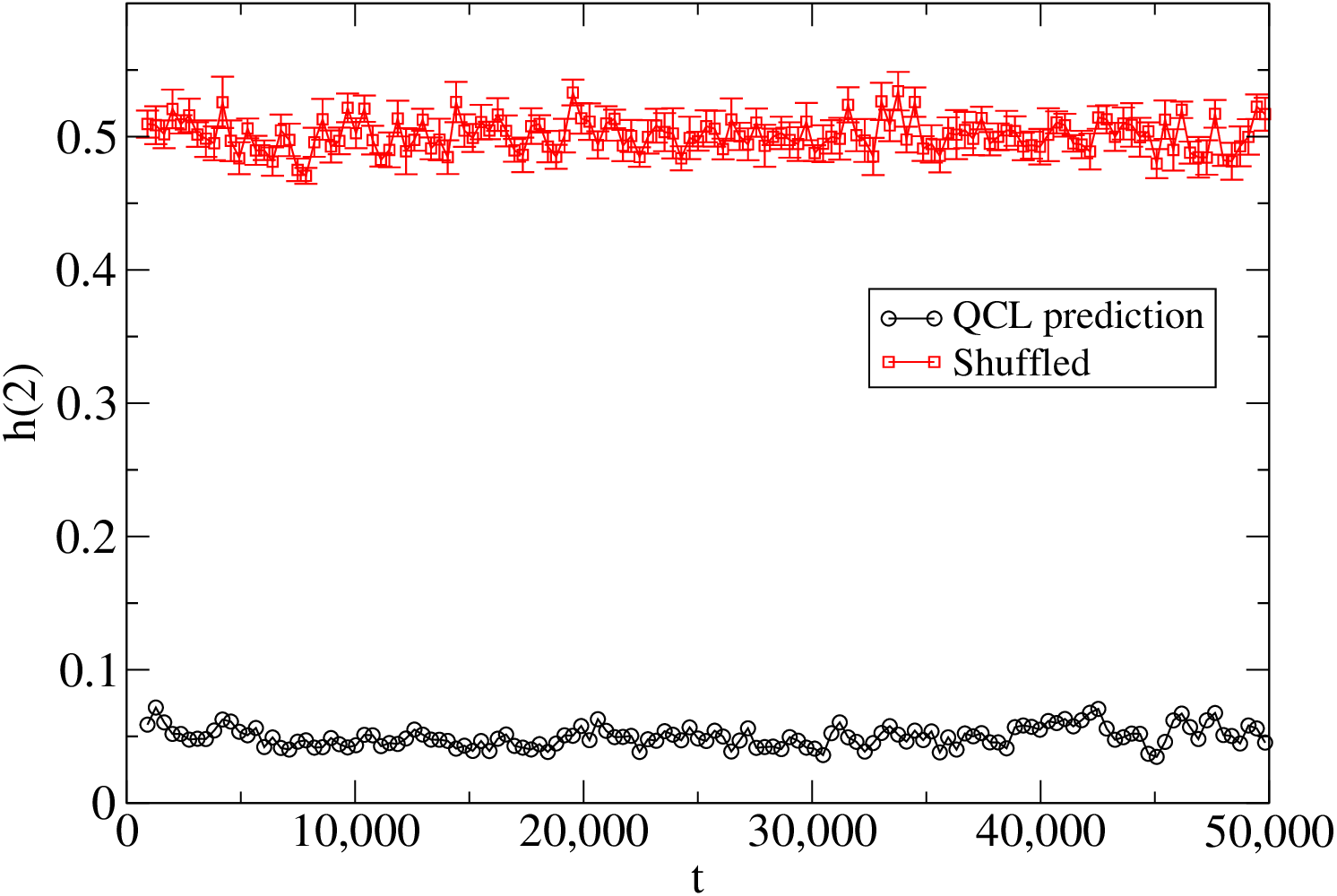}
\caption{Hurst
 exponent $h(2)$ of the IAR time series and those of the shuffled time series.
For QCL predictions, the typical magnitude of the fitting error is on the order of the symbol size or smaller.}
\label{fig19}
\end{figure}
\begin{figure}[H]
\vspace{1.5cm}
\isPreprints{\centering}{} 
\includegraphics[width=9.0 cm]{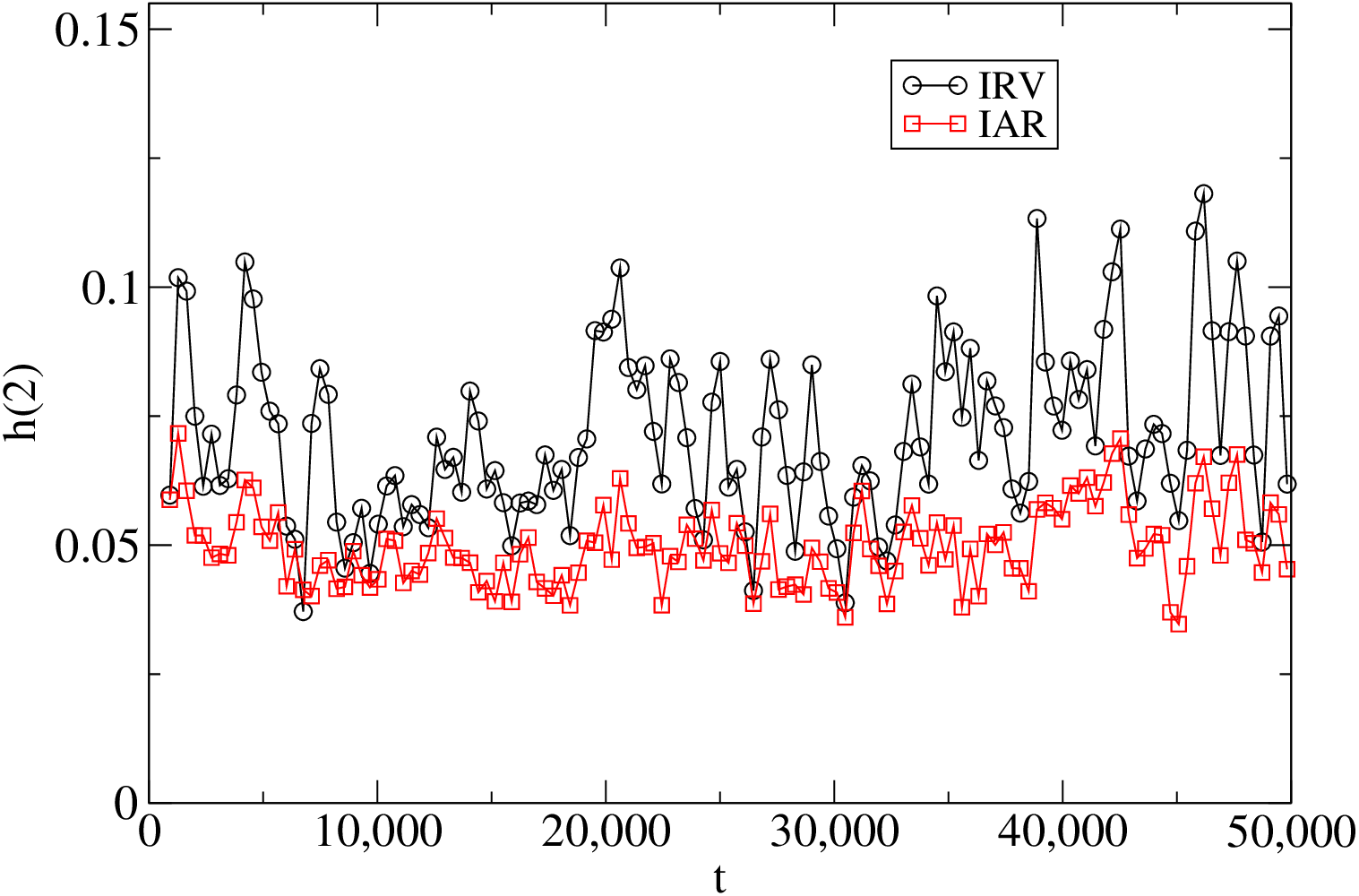}
\caption{Comparison
 of the Hurst exponent $h(2)$ computed for the IRV and IAR time series.
The typical magnitude of the fitting error is on the order of the symbol size or smaller.}\label{fig20}
\end{figure}

Figure \ref{fig19} plots the Hurst exponent of the increment series $\mathrm{IAR}$. The estimated values fall within the range of $h(2) \approx 0.04$--$0.07$ (on average $\sim$0.05), indicating anti-persistent behavior. These values are smaller than those obtained for the IRV (see also Figure \ref{fig20}). 
 Prior studies~\cite{garcin2022long,takaishi2025multifractality} examining RV constructed with varying sampling frequencies report that the estimated Hurst exponent decreases as the number of intraday observations used in RV construction becomes smaller. The present results are consistent with this \mbox{empirical regularity.}

\vspace{-9pt}

\begin{figure}[H]
\vspace{1.5cm}
\isPreprints{\centering}{} 
\includegraphics[width=9.0 cm]{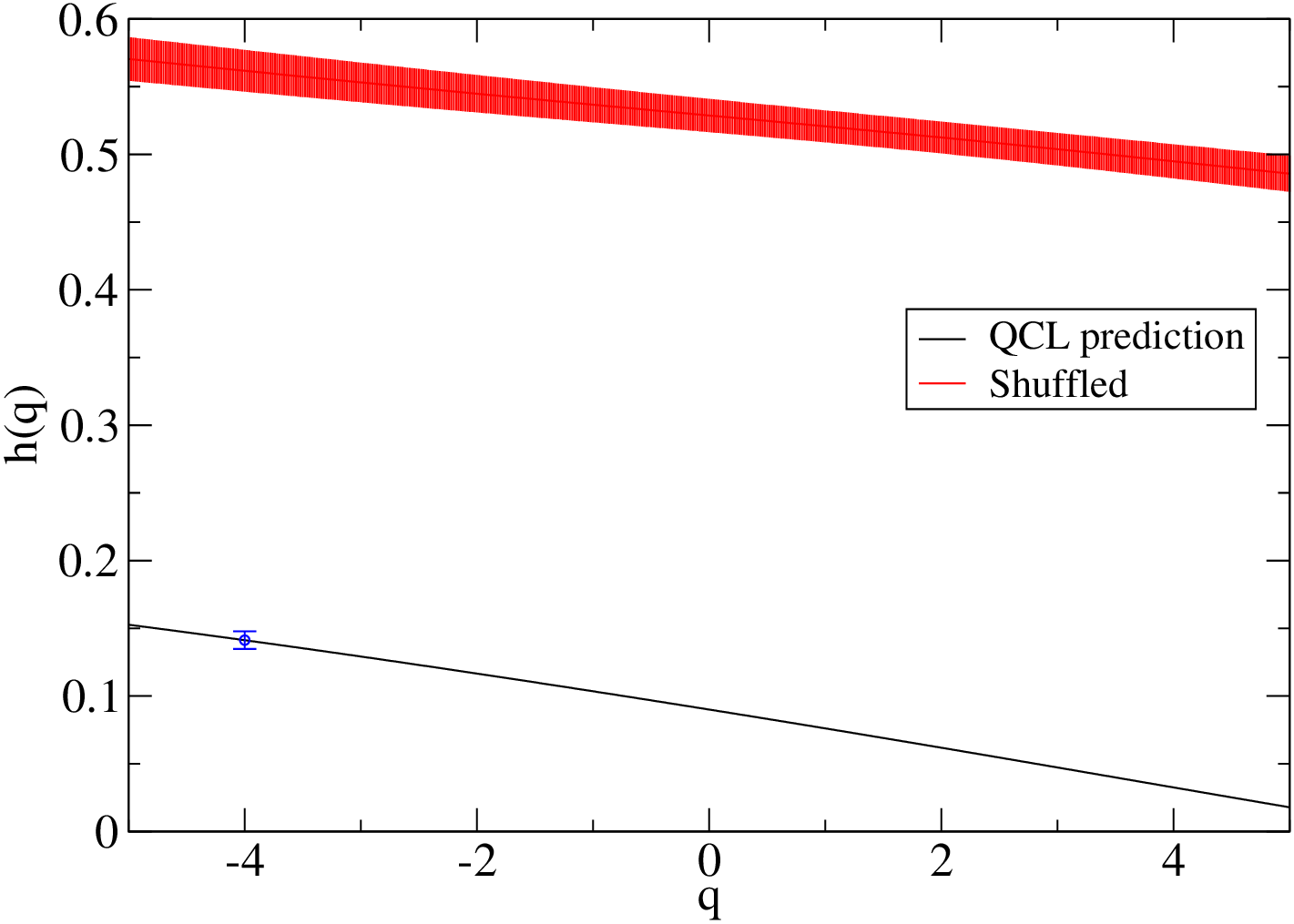}
\caption{$h(q)$ of 
 the IRV time series in a representative window.
The blue symbol with error bars in the figure indicates the typical magnitude of the fitting error.}\label{fig21}
\end{figure}

Next, we investigate the multifractal properties of the IRV series. Figure \ref{fig21} 
 displays the generalized Hurst exponent $h(q)$ for both the QCL-predicted IRV series and its shuffled counterpart within a representative window. The IRV series exhibits a monotonically decreasing $h(q)$ as $q$ increases, demonstrating clear multifractality. Although the shuffled series shows $h(2)$ values shifted toward $0.5$, consistent with random behavior, its $h(q)$ still decreases monotonically with $q$, indicating that multifractality persists even after temporal correlations are removed. Comparing the variation range of $h(q)$ between the IRV and shuffled series reveals that the width of the $h(q)$ spectrum remains largely unchanged.

\begin{figure}[H]
\vspace{1.5cm}
\isPreprints{\centering}{} 
\includegraphics[width=9.0 cm]{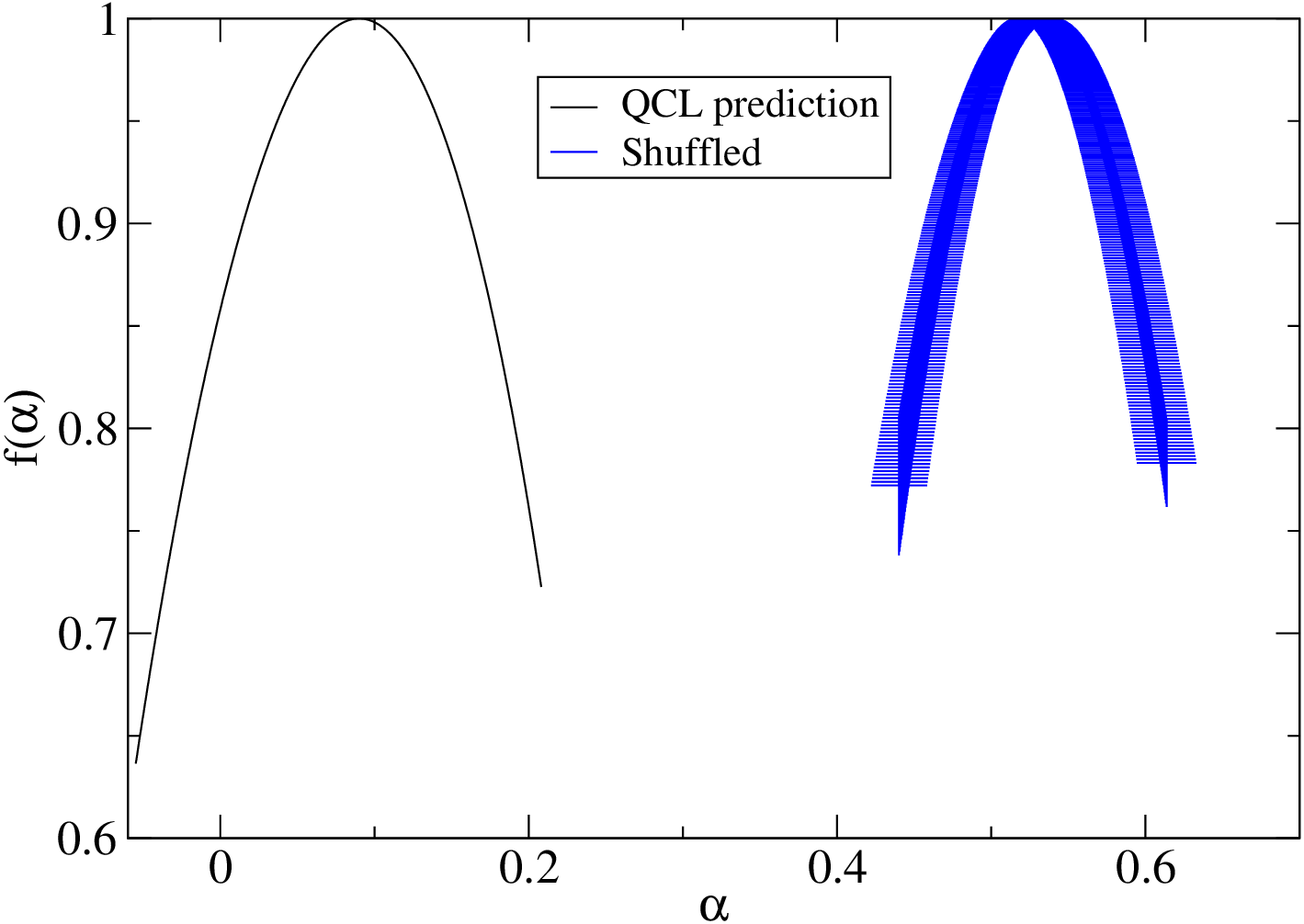}
\caption{$f(\alpha)$ of the IRV time series in a representative window.}\label{fig22}
\end{figure}

Figure \ref{fig22} presents the singularity spectra $f(\alpha)$ computed for the IRV time series and its shuffled counterpart within a representative window. Both $f(\alpha)$ exhibit a finite width in $\alpha$, indicating
that each series possesses multifractal characteristics. 

\begin{figure}[H]
\vspace{1.5cm}
\isPreprints{\centering}{} 
\includegraphics[width=9.0 cm]{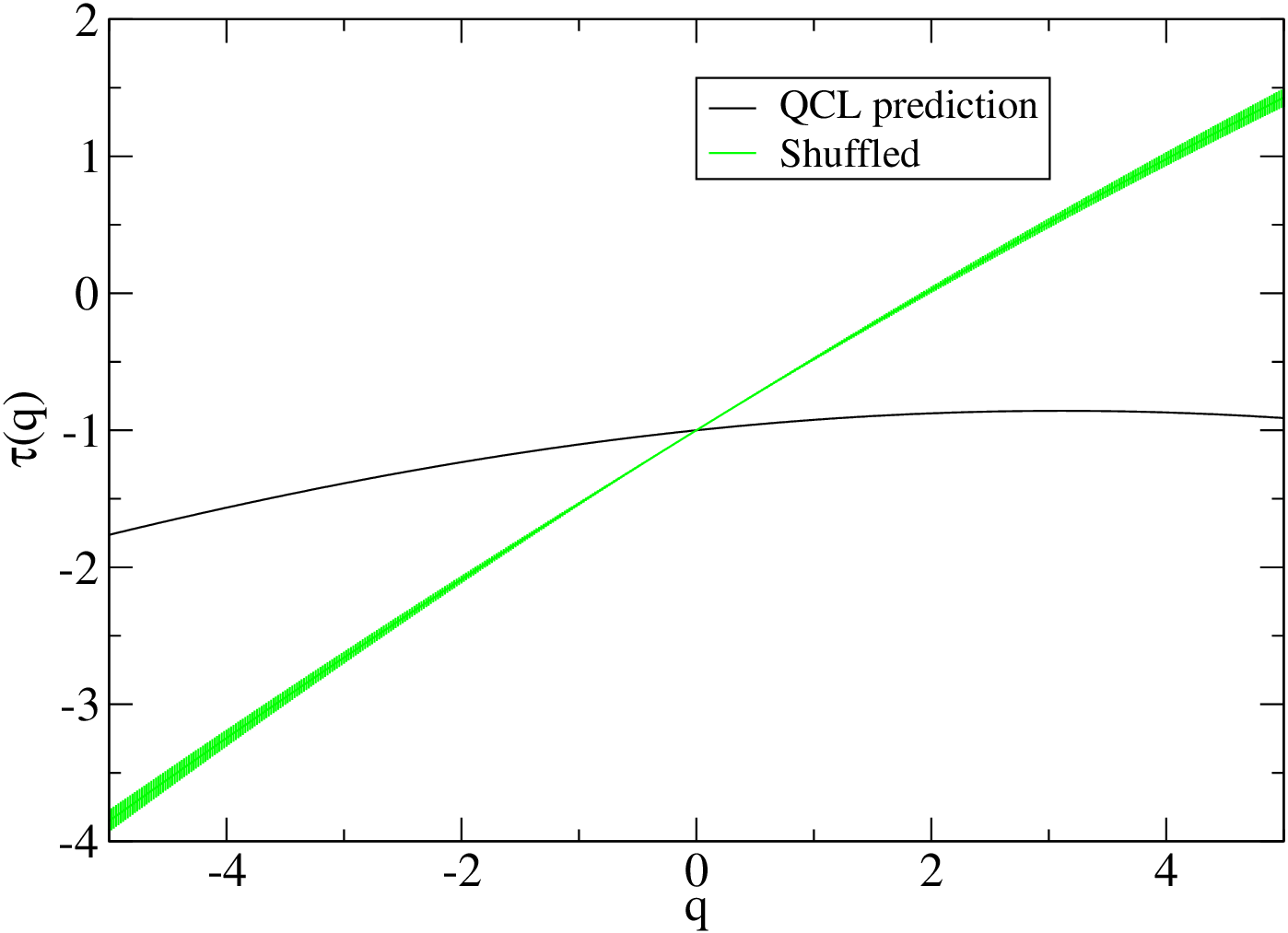}
\caption{$\tau(q)$ of
 the IRV time series in a representative window.}\label{fig23}
\end{figure}

A similar pattern is observed for $\tau(q)$: in both the original and shuffled series, $\tau(q)$ remains nonlinear, further confirming the presence of multifractality (Figure \ref{fig23}).
After shuffling, the position of the peak in $f(\alpha)$ shifts. This is because the value of $h(0)$, which corresponds to the peak position, increases after the shuffling process. Similarly, the slope of $\tau(q)$ becomes steeper after shuffling; this is attributed to the fact that the values of $h(q)$, which approximately represent the magnitude of the slope of $\tau(q)$, increase after shuffling, as shown in Figure \ref{fig23}.

\vspace{-9pt}

\vspace{-6pt}

\begin{figure}[H]
\vspace{1.5cm}
\isPreprints{\centering}{} 
\includegraphics[width=9.0 cm]{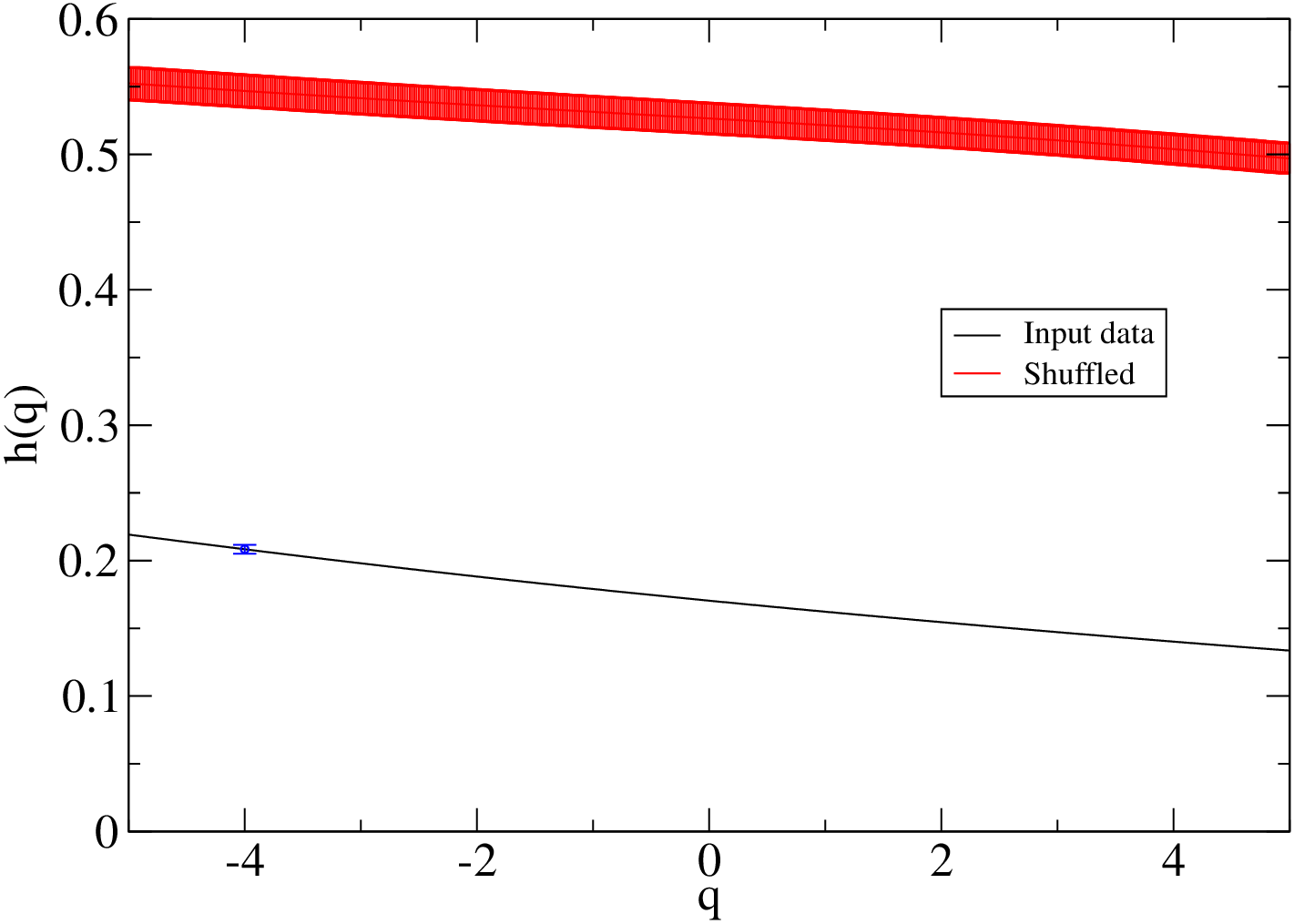}
\caption{$h(q)$ of 
 the IRV time series computed from the input data
The blue symbol with error bars in the figure indicates the typical magnitude of the fitting error.}\label{fig24}
\end{figure}

Figures \ref{fig24}--26
 present the multifractal analysis results of $h(q)$, $f(\alpha)$, and $\tau(q)$ for the IRV time series constructed from the RV data used as input to the QCL model. The input data exhibit multifractal behavior similar to that observed in the QCL-generated time series (Figures \ref{fig21}--\ref{fig23}), demonstrating that the multifractal characteristics of the empirical IRV series are similar to those generated by the QCL model.

Overall, our results demonstrate that the QCL-generated time series qualitatively reproduces empirical properties of Bitcoin volatility, including anti-persistence and multifractality in IRV.

\begin{figure}[H]
\vspace{1.5cm}
\isPreprints{\centering}{} 
\includegraphics[width=9.0 cm]{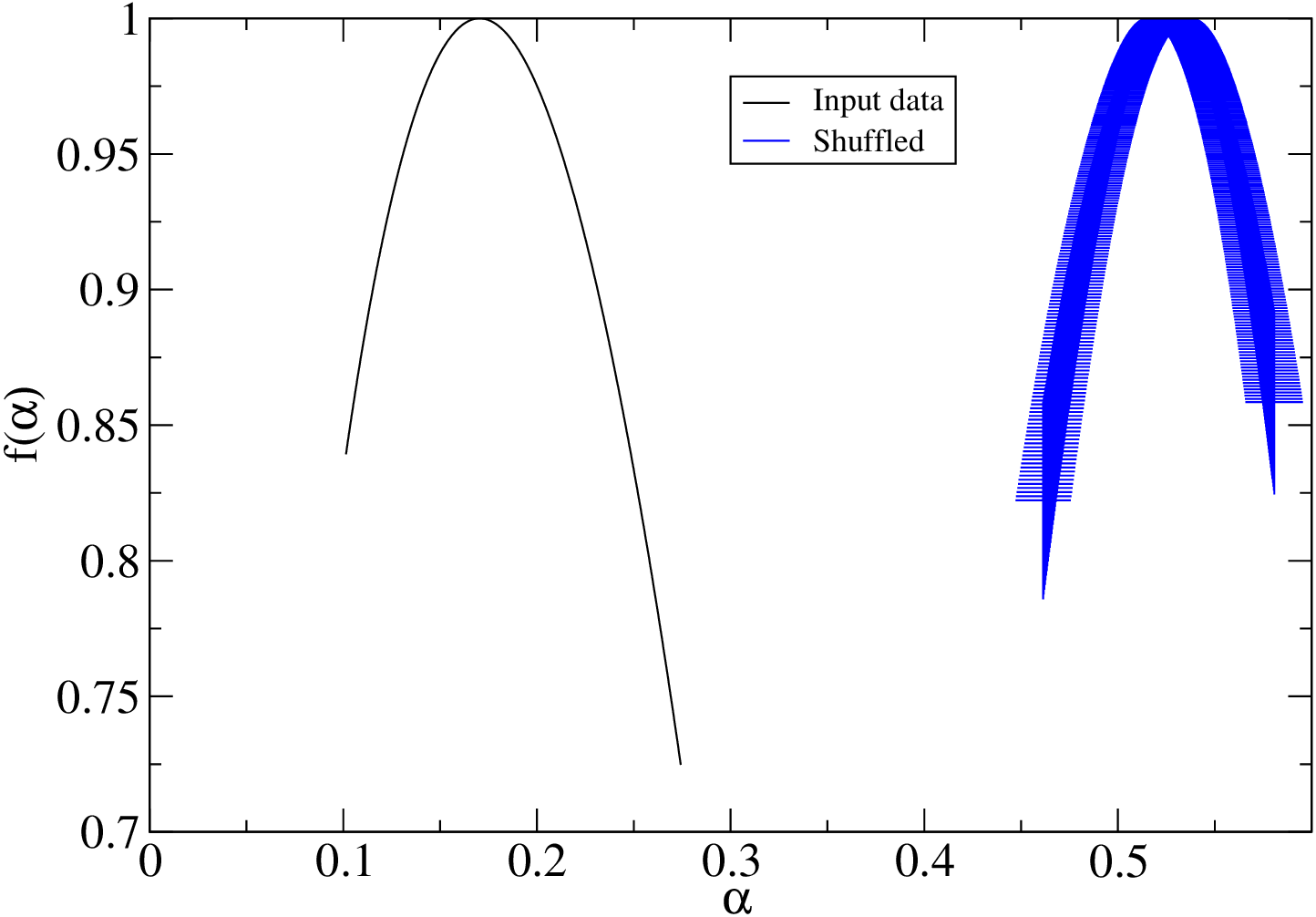}
\caption{$f(\alpha)$ of the IRV time series  computed from the input data}\label{fig25}
\end{figure}   
\vspace{-9pt}
\begin{figure}[H]
\vspace{1.5cm}
\isPreprints{\centering}{} 
\includegraphics[width=9.0 cm]{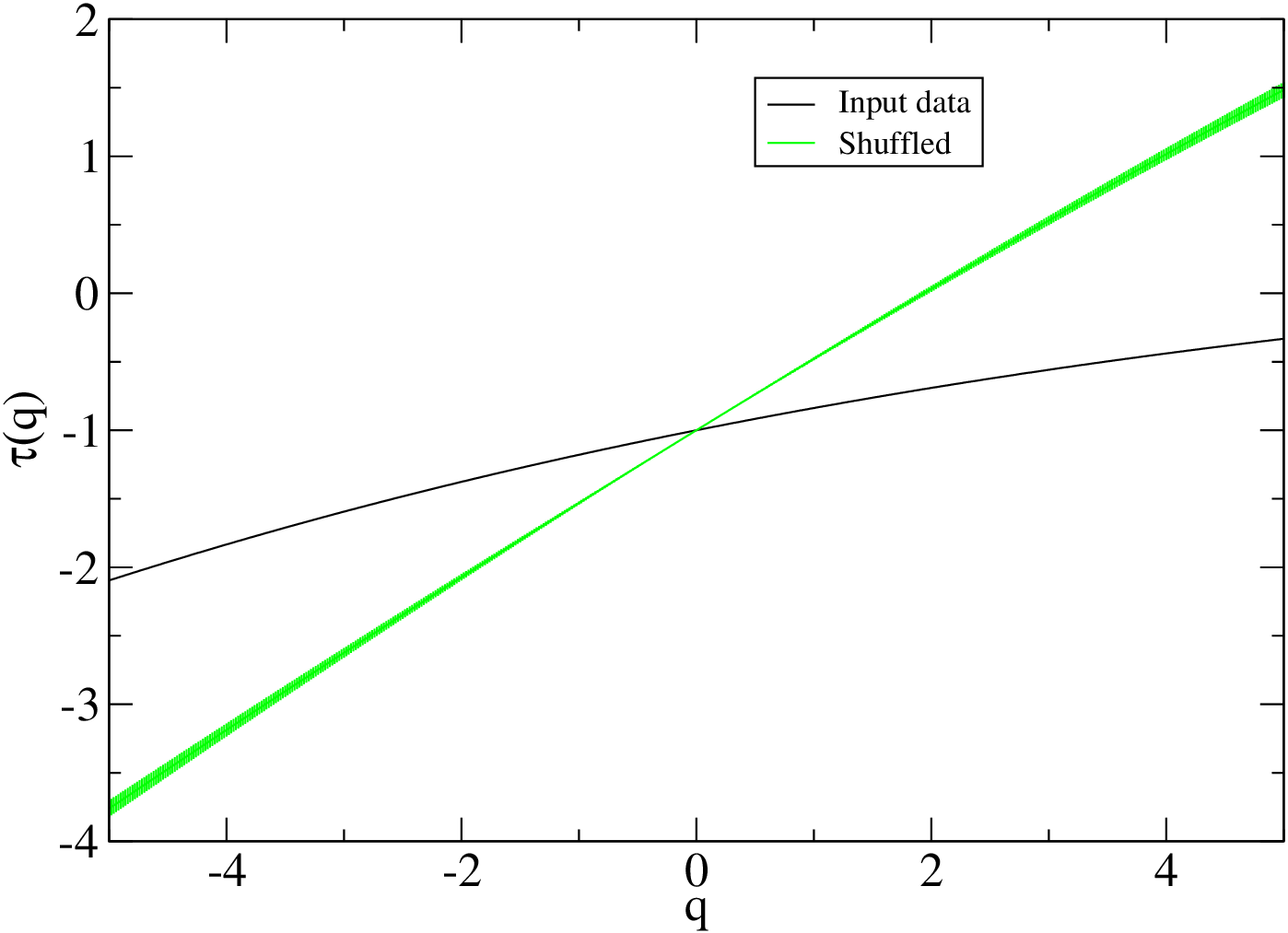}
\caption{$\tau(q)$ of 
 the IRV time series computed from the input data}\label{fig26}
\end{figure}   

\section{Discussion and Conclusions}\label{sec7}
 
In this study, the QCL was applied to approximate
the RV time series of Bitcoin, and the
resulting predicted series is subjected to multifractal analysis. We discuss the
significance and interpretation of the results below.
 
\subsection{Effectiveness of QCL for Volatility Approximation
}

As shown in Figure \ref{fig4}, the optimized quantum circuit reproduces the
broad temporal variation in the empirical RV series. Unlike GARCH-type
models, which require a pre-specified functional form, the present
approach directly approximates the volatility function from data,
eliminating the need for model selection. A limitation, however, is
that the single-qubit, three-parameter circuit employed here has
limited expressive power; deeper multi-qubit circuits or the data re-uploading technique~\cite{perez2020data,perez2021one} 
{may improve the \mbox{approximation accuracy.} }

\subsection{Multifractality of the Return Time Series}
 Multifractality in $h(q)$ is observed for the predicted return series
(Figure \ref{fig9}), and partial multifractality persists even after shuffling
(Figures \ref{fig9}--\ref{fig11}). However, analysis based on finite-length Gaussian random time series (Figures \ref{fig12}--\ref{fig14}) indicates the presence of multifractality arising from finite-size effects, suggesting that the partial multifractality observed after shuffling can be attributed to such finite-size effects.
Analogous results are obtained for the original data (Figures \ref{fig15}--\ref{fig17}),
confirming that the QCL model reproduces the multifractal
structure of the empirical return time series. 
 
\subsection{Anti-Persistence of IRV}
The IRV series exhibits $h(2) \approx 0.05$--$0.1$ (Figure \ref{fig18}), which is
consistent with the rough volatility findings of Gatheral et
al.~\cite{gatheral2018volatility}. This anti-persistence is reproduced by the
QCL model, and the reversion of $h(2)$ to approximately
$0.5$ after shuffling (Figure \ref{fig18}) clearly attributes the
anti-persistence to temporal correlations. The smaller $h(2)$ values
of the IAR time series relative to IRV (Figure \ref{fig20}) are consistent with the
finding that the RV from fewer intraday samples produces lower Hurst exponents~\cite{takaishi2025multifractality}.
{The randomness observed in the return series is expected from the model definition; however, the emergence of anti-persistence in volatility is a non-trivial issue. The fact that the volatility time series predicted by the QCL model exhibits anti-persistence can be regarded as evidence that the model successfully captures the essential characteristics of volatility dynamics in real data. }

\subsection{Comparison and Future Directions}
The QCL model qualitatively reproduces the major multifractal
properties of the empirical Bitcoin RV series: near-random return
dynamics ($h(2) \approx 0.5$), multifractality of returns, and
anti-persistence of IRV ($h(2) \approx 0.05$--$0.1$). 

It is noted that the present model adopts a single-qubit parameterized quantum circuit with a highly simplified structure, inspired by classical parsimonious models such as GARCH(1, 1), which successfully capture some stylized facts of financial markets using only a small number of parameters. The primary objective of this study is to examine whether such a minimal QCL framework can reproduce the essential statistical properties of financial time series. However, in doing so, it may constrain the expressive power of the model.
 Therefore, extensions to deeper architectures and multi-qubit circuits are expected to be necessary for enhancing representational capacity, which we leave for future investigations.
Furthermore, demonstrating the practical utility of the proposed approach requires systematic benchmarking against established models, including GARCH, HAR, and Realized GARCH, as well as evaluation in tasks such as volatility forecasting. Such analyses are beyond the scope of the present study and are reserved for future work. Accordingly, this study should be regarded as an initial step toward assessing the capability of QCL to capture statistical properties of financial time series, while comprehensive performance comparisons and applications to forecasting and risk management remain important directions for future research.

Additional future directions include the confirmation of empirical facts
observed in financial asset prices---such as the Taylor
effect~\cite{ding1996modeling,haas2009persistence,takaishi2018taylor}, the leverage
effect~\cite{bouchaud2001leverage,qiu2006return,shen2009return}, the
Zumbach
effect~\cite{zumbach2003volatility,zumbach2009time}, and the
inverse cubic law in return
distributions~\cite{gopikrishnan1998inverse}---as well as applications to
other asset classes (equities, foreign exchange, and commodities) to
assess the generality of the approach.






\funding{This research was funded by JSPS KAKENHI}, Grant Number 26K04856.

\dataavailability{The 
 original data presented in the study are openly available at
 [\url{https://github.com/takaishi-physics/QCL-volatility}], accessed on 23 June 2026.}

\acknowledgments{Numerical calculations for this work were carried out at the facilities of the Institute of Statistical Mathematics.}

\conflictsofinterest{The author declares no conflicts of interest. }




\isPreprints{}{
} 


\reftitle{References}



\begin{thebibliography}{99}

\bibitem[Cont(2001)]{Cont2001QF}
Cont, R. Empirical Properties of Asset Returns: Stylized Facts and Statistical Issues. \emph{Quant. Financ.} \textbf{2001}, \emph{1}, 223--236.

\bibitem[Engle(1982)]{Engle1982autoregressive}
Engle, R.F. Autoregressive conditional heteroscedasticity with estimates of the variance of United Kingdom inflation. \emph{Econom. J. Econom. Soc.} \textbf{1982}, 987--1007.

\bibitem[Bollerslev(1986)]{Bollerslev1986JOE}
Bollerslev, T. Generalized Autoregressive Conditional Heteroskedasticity. \emph{J. Econom.} \textbf{1986}, \emph{31}, 307--327.

\bibitem[Black(1976)]{Black1976}
Black, F. Studies of Stock Market Volatility Changes. In \emph{1976 Proceedings of the  Business and Economic Statistics Section}; American Statistical Association: Alexandria, VA, USA, 
 1976; pp. 177--181.

\bibitem[Christie(1982)]{Christie1982stochastic}
Christie, A.A. The stochastic behavior of common stock variances: Value, leverage and interest rate effects. \emph{J. Financ. Econ.} \textbf{1982}, \emph{10}, 407--432.

\bibitem[Nelson(1991)]{Nelson1991Econ}
Nelson, D. Conditional Heteroskedasticity in Asset Returns: A New Approach. \emph{Econometrica} \textbf{1991}, \emph{59}, 347--370.

\bibitem[Glosten et~al.(1993)Glosten, Jaganathan, and Runkle]{Glosten1993JOF}
Glosten, L.; Jaganathan, R.; Runkle, D. On the Relation Between the Expected Value and the Volatility of the Nominal Excess on Stocks. \emph{J. Financ.} \textbf{1993}, \emph{48}, 1779--1801.

\bibitem[Sentana(1995)]{Sentana1995RES}
Sentana, E. Quadratic ARCH Models. \emph{Rev. Econ. Stud.} \textbf{1995}, \emph{62}, 639--661.

\bibitem[Takaishi(2017)]{takaishi2017rational}
Takaishi, T. Rational GARCH model: An empirical test for stock returns. \emph{Phys. A} \textbf{2017}, \emph{473}, 451--460.

\bibitem[Takaishi(2018)]{takaishi2018volatility}
Takaishi, T. Volatility estimation using a rational GARCH model. \emph{Quant. Financ. Econ.} \textbf{2018}, \emph{2}, 127--36.

\bibitem[Ding et~al.(1993)Ding, Granger, and Engle]{ding1993long}
Ding, Z.; Granger, C.W.; Engle, R.F. A long memory property of stock market returns and a new model. \emph{J. Empir. Financ.} \textbf{1993}, \emph{1}, 83--106.

\bibitem[Taylor(1982)]{taylor1982}
Taylor, S.J. \emph{Financial Returns Modelled by the Product of Two Stochastic Processes, a Study of Daily Sugar Prices 1961--79}; 
In \emph{Time Series Analysis: Theory and Practice};
North-Holland:,
Amsterdam, 1982; pp. 203--226.
 

\bibitem[Taylor(1986)]{taylor1986modelling}
Taylor, S.J. \emph{Modelling Financial Time Series}; John Wiley \& New Jersey, 1986.

\bibitem[Takaishi(2006/10)]{Takaishi2006/10}
Takaishi, T. Bayesian estimation of GARCH model by hybrid Monte Carlo. In \emph{Proceedings of the  9th Joint International Conference on Information Sciences (JCIS-06)}; Atlantis Press: Dordrecht, The Netherlands, 
 2006; pp. 661--664. \url{https://doi.org/10.2991/jcis.2006.159}.

\bibitem[Kim et~al.(1998)Kim, Shephard, and Chib]{kim1998stochastic}
Kim, S.; Shephard, N.; Chib, S. Stochastic volatility: Likelihood inference and comparison with ARCH models. \emph{Rev. Econ. Stud.} \textbf{1998}, \emph{65}, 361--393.

\bibitem[Jacquier et~al.(2002)Jacquier, Polson, and Rossi]{jacquier2002bayesian}
Jacquier, E.; Polson, N.G.; Rossi, P.E. Bayesian analysis of stochastic volatility models. \emph{J. Bus. Econ. Stat.} \textbf{2002}, \emph{20}, 69--87.

\bibitem[Omori et~al.(2007)Omori, Chib, Shephard, and Nakajima]{omori2007stochastic}
Omori, Y.; Chib, S.; Shephard, N.; Nakajima, J. Stochastic volatility with leverage: Fast and efficient likelihood inference. \emph{J. Econom.} \textbf{2007}, \emph{140}, 425--449.

\bibitem[Gupta(1998)]{gupta1998introduction}
Gupta, R. Introduction to lattice QCD. \emph{arXiv} \textbf{1998}, arXiv:hep-lat/9807028.

\bibitem[Lippert(2007)]{lippert2007hybrid}
Lippert, T. The hybrid Monte Carlo algorithm for quantum chromodynamics. In \emph{Field Theoretical Tools for Polymer and Particle Physics}; Springer: Berlin/Heidelberg, Germany, 2007; pp. 122--132.

\bibitem[Duane et~al.(1987)Duane, Kennedy, Pendleton, and Roweth]{duane1987hybrid}
Duane, S.; Kennedy, A.D.; Pendleton, B.J.; Roweth, D. Hybrid Monte Carlo. \emph{Phys. Lett. B} \textbf{1987}, \emph{195}, 216--222.

\bibitem[Sexton and Weingarten(1992)]{sexton1992hamiltonian}
Sexton, J.C.; Weingarten, D.H. Hamiltonian evolution for the hybrid Monte Carlo algorithm. \emph{Nucl. Phys. B} \textbf{1992}, \emph{380}, 665--677.

\bibitem[de~Forcrand and Takaishi(1997)]{Takaishi1997Fast}
de Forcrand, P.; Takaishi, T. Fast fermion Monte Carlo. \emph{Nucl. Phys. B-Proc. Suppl.} \textbf{1997}, \emph{53}, 968--970.

\bibitem[Hasenbusch(2001)]{hasenbusch2001speeding}
Hasenbusch, M. Speeding up the hybrid Monte Carlo algorithm for dynamical fermions. \emph{Phys. Lett. B} \textbf{2001}, \emph{519}, 177--182.

\bibitem[Takaishi(2000)]{takaishi2000choice}
Takaishi, T. Choice of integrator in the hybrid Monte Carlo algorithm. \emph{Comput. Phys. Commun.} \textbf{2000}, \emph{133}, 6--17.

\bibitem[Takaishi and de~Forcrand(2002)]{takaishi2002odd}
Takaishi, T.; de Forcrand, P. Odd-flavor Hybrid Monte Carlo Algorithm for Lattice QCD. \emph{Int. J. Mod. Phys. C} \textbf{2002}, \emph{13}, 343--365.

\bibitem[Clark(2006)]{clark2006rational}
Clark, M.A. The rational hybrid Monte Carlo algorithm. \emph{arXiv} \textbf{2006}, \emph{LAT2006}, 004.
 arXiv:hep-lat/0610048.



\bibitem[Clark and Kennedy(2007)]{clark2007accelerating}
Clark, M.; Kennedy, A. Accelerating dynamical-fermion computations using the Rational Hybrid Monte Carlo algorithm with multiple pseudofermion fields. \emph{Phys. Rev. Lett.} \textbf{2007}, \emph{98}, 051601.

\bibitem[Takaishi(2009)]{takaishi2009bayesian}
Takaishi, T. Bayesian Inference of stochastic volatility model by Hybrid Monte Carlo. \emph{J. Circuits Syst. Comput.} \textbf{2009}, \emph{18}, 1381--1396.

\bibitem[Takaishi(2014)]{takaishi2014RSV}
Takaishi, T. Bayesian estimation of realized stochastic volatility model by Hybrid Monte Carlo algorithm. \emph{J. Phys. Conf. Ser.} \textbf{2014}, \emph{490}, 012092.

\bibitem[Andersen et~al.(2003)Andersen, Bollerslev, Diebold, and Labys]{andersen2003modeling}
Andersen, T.G.; Bollerslev, T.; Diebold, F.X.; Labys, P. Modeling and forecasting realized volatility. \emph{Econometrica} \textbf{2003}, \emph{71}, 579--625.

\bibitem[Granger and Joyeux(1980)]{granger1980introduction}
Granger, C.W.; Joyeux, R. An introduction to long-memory time series models and fractional differencing. \emph{J. Time Ser. Anal.} \textbf{1980}, \emph{1}, 15--29.

\bibitem[Baillie et~al.(1996)Baillie, Bollerslev, and Mikkelsen]{baillie1996fractionally}
Baillie, R.T.; Bollerslev, T.; Mikkelsen, H.O. Fractionally integrated generalized autoregressive conditional heteroskedasticity. \emph{J. Econom.} \textbf{1996}, \emph{74}, 3--30.

\bibitem[Tayefi and Ramanathan(2012)]{tayefi2012overview}
Tayefi, M.; Ramanathan, T. An overview of FIGARCH and related time series models. \emph{Austrian J. Stat.} \textbf{2012}, \emph{41}, 175--196.

\bibitem[Gatheral et~al.(2018)Gatheral, Jaisson, and Rosenbaum]{gatheral2018volatility}
Gatheral, J.; Jaisson, T.; Rosenbaum, M. Volatility is rough. \emph{Quant. Financ.} \textbf{2018}, \emph{18}, 933--949.

\bibitem[Bennedsen et~al.(2022)Bennedsen, Lunde, and Pakkanen]{bennedsen2022decoupling}
Bennedsen, M.; Lunde, A.; Pakkanen, M.S. Decoupling the short-and long-term behavior of stochastic volatility. \emph{J. Financ. Econom.} \textbf{2022}, \emph{20}, 961--1006.

\bibitem[Livieri et~al.(2018)Livieri, Mouti, Pallavicini, and Rosenbaum]{livieri2018rough}
Livieri, G.; Mouti, S.; Pallavicini, A.; Rosenbaum, M. Rough volatility: Evidence from option prices. \emph{IISE Trans.} \textbf{2018}, \emph{50}, 767--776.

\bibitem[Floc'h(2022)]{floc2022roughness}
Floc’h, F.L. Roughness of the Implied Volatility. \emph{arXiv} \textbf{2022}, arXiv:2207.04930.

\bibitem[Takaishi(2025)]{takaishi2025multifractality}
Takaishi, T. Multifractality and sample size influence on Bitcoin volatility patterns. \emph{Financ. Res. Lett.} \textbf{2025}, \emph{74}, 106683.

\bibitem[Bariviera et~al.(2023)Bariviera, Fabregat-Aibar, and Sorrosal-Forradellas]{bariviera2023disentangling}
Bariviera, A.F.; Fabregat-Aibar, L.; Sorrosal-Forradellas, M.T. Disentangling the impact of economic and health crises on financial markets. \emph{Res. Int. Bus. Financ.} \textbf{2023}, \emph{65}, 101928.

\bibitem[Takaishi(2025)]{takaishi2025impact}
Takaishi, T. Impact of the COVID-19 pandemic on the financial market efficiency of price returns, absolute returns, and volatility increment: Evidence from stock and cryptocurrency markets. \emph{J. Risk Financ. Manag.} \textbf{2025}, \emph{18}, 237.

\bibitem[Clark(1973)]{clark1973subordinated}
Clark, P.K. A subordinated stochastic process model with finite variance for speculative prices. \emph{Econometrica} \textbf{1973}, \emph{41}, 135--155.

\bibitem[Tauchen and Pitts(1983)]{tauchen1983price}
Tauchen, G.E.; Pitts, M. The price variability-volume relationship on speculative markets. \emph{Econom. J. Econom. Soc.} \textbf{1983}, 485--505.

\bibitem[Andersen(1996)]{andersen1996return}
Andersen, T.G. Return volatility and trading volume: An information flow interpretation of stochastic volatility. \emph{J. Financ.} \textbf{1996}, \emph{51}, 169--204.

\bibitem[Lamoureux and Lastrapes(1990)]{lamoureux1990heteroskedasticity}
Lamoureux, C.G.; Lastrapes, W.D. Heteroskedasticity in stock return data: Volume versus GARCH effects. \emph{J. Financ.} \textbf{1990}, \emph{45}, 221--229.

\bibitem[Sharma et~al.(1996)Sharma, Mougoue, and Kamath]{sharma1996heteroscedasticity}
Sharma, J.L.; Mougoue, M.; Kamath, R. Heteroscedasticity in stock market indicator return data: Volume versus GARCH effects. \emph{Appl. Financ. Econ.} \textbf{1996}, \emph{6}, 337--342.

\bibitem[Miyakoshi(2002)]{miyakoshi2002arch}
Miyakoshi, T. ARCH versus information-based variances: Evidence from the Tokyo stock market. \emph{Jpn. World Econ.} \textbf{2002}, \emph{14}, 215--231.

\bibitem[Bose and Rahman(2015)]{bose2015examining}
Bose, S.; Rahman, H. Examining the relationship between stock return volatility and trading volume: New evidence from an emerging economy. \emph{Appl. Econ.} \textbf{2015}, \emph{47}, 1899--1908.

\bibitem[Takaishi and Chen(2016)]{takaishi2016relationship}
Takaishi, T.; Chen, T.T. The relationship between trading volumes, number of transactions, and stock volatility in GARCH models. \emph{J. Phys. Conf. Ser.} \textbf{2016}, \emph{738}, 012097.

\bibitem[Takaishi(2020)]{takaishi2022Hurst}
Takaishi, T. Hurst exponent and Multifractal Properties in the Time Series of Bitcoin Trading Volume. \emph{Int. J. Eng. Res. Appl.} \textbf{2020}, \emph{12}, 24--29.

\bibitem[Andersen and Bollerslev(1998)]{andersen1998answering}
Andersen, T.G.; Bollerslev, T. Answering the skeptics: Yes, standard volatility models do provide accurate forecasts. \emph{Int. Econ. Rev.} \textbf{1998}, \emph{39}, 885--905.

\bibitem[Barndorff-Nielsen and Shephard(2002)]{barndorff2002econometric}
Barndorff-Nielsen, O.E.; Shephard, N. Econometric analysis of realized volatility and its use in estimating stochastic volatility models. \emph{J. R. Stat. Soc. Ser. B Stat. Methodol.} \textbf{2002}, \emph{64}, 253--280.

\bibitem[McAleer and Medeiros(2008)]{mcaleer2008realized}
McAleer, M.; Medeiros, M.C. Realized volatility: A review. \emph{Econom. Rev.} \textbf{2008}, \emph{27}, 10--45.

\bibitem[Hansen et~al.(2012)Hansen, Huang, and Shek]{hansen2012realized}
Hansen, P.R.; Huang, Z.; Shek, H.H. Realized GARCH: A joint model for returns and realized measures of volatility. \emph{J. Appl. Econom.} \textbf{2012}, \emph{27}, 877--906.

\bibitem[Hansen and Huang(2016)]{hansen2016exponential}
Hansen, P.R.; Huang, Z. Exponential GARCH modeling with realized measures of volatility. \emph{J. Bus. Econ. Stat.} \textbf{2016}, \emph{34}, 269--287.

\bibitem[Takahashi et~al.(2009)Takahashi, Omori, and Watanabe]{takahashi2009estimating}
Takahashi, M.; Omori, Y.; Watanabe, T. Estimating stochastic volatility models using daily returns and realized volatility simultaneously. \emph{Comput. Stat. Data Anal.} \textbf{2009}, \emph{53}, 2404--2426.

\bibitem[Koopman and Scharth(2013)]{koopman2013analysis}
Koopman, S.J.; Scharth, M. The analysis of stochastic volatility in the presence of daily realized measures. \emph{J. Financ. Econom.} \textbf{2013}, \emph{11}, 76--115.

\bibitem[Takaishi(2018)]{takaishi2018bias}
Takaishi, T. Bias correction in the realized stochastic volatility model for daily volatility on the Tokyo Stock Exchange. \emph{Phys. A} \textbf{2018}, \emph{500}, 139--154.

\bibitem[Corsi(2009)]{corsi2009simple}
Corsi, F. A simple approximate long-memory model of realized volatility. \emph{J. Financ. Econom.} \textbf{2009}, \emph{7}, 174--196.

\bibitem[Huang et~al.(2016)Huang, Liu, and Wang]{huang2016modeling}
Huang, Z.; Liu, H.; Wang, T. Modeling long memory volatility using realized measures of volatility: A realized HAR GARCH model. \emph{Econ. Model.} \textbf{2016}, \emph{52}, 812--821.

\bibitem[Mitarai et~al.(2018)Mitarai, Negoro, Kitagawa, and Fujii]{mitarai2018quantum}
Mitarai, K.; Negoro, M.; Kitagawa, M.; Fujii, K. Quantum circuit learning. \emph{Phys. Rev. A} \textbf{2018}, \emph{98}, 032309.

\bibitem[Bauer et~al.(2020)Bauer, Bravyi, Motta, and Chan]{bauer2020quantum}
Bauer, B.; Bravyi, S.; Motta, M.; Chan, G.K.L. Quantum algorithms for quantum chemistry and quantum materials science. \emph{Chem. Rev.} \textbf{2020}, \emph{120}, 12685--12717.

\bibitem[Takaishi(2025)]{takaishi2025volatility}
Takaishi, T. Volatility time series modeling by single-qubit quantum circuit learning. \emph{arXiv} \textbf{2025}, arXiv:2512.10584.

\bibitem[Kantelhardt et~al.(2002)Kantelhardt, Zschiegner, Koscielny-Bunde, Havlin, Bunde, and Stanley]{kantelhardt2002multifractal}
Kantelhardt, J.W.; Zschiegner, S.A.; Koscielny-Bunde, E.; Havlin, S.; Bunde, A.; Stanley, H.E. Multifractal detrended fluctuation analysis of nonstationary time series. \emph{Phys. A} \textbf{2002}, \emph{316}, 87--114.

\bibitem[Nuyts and Platten(2001)]{Nuyts2001Physica}
Nuyts, J.; Platten, I. Phenomenology of the term structure of interest rates with Padé Approximants. \emph{Phys. A} \textbf{2001}, \emph{299}, 528--546.

\bibitem[Chen and Takaishi(2013)]{chen2013empirical}
Chen, T.T.; Takaishi, T. Empirical study of the GARCH model with rational errors. \emph{J. Phys. Conf. Ser.} \textbf{2013}, \emph{454}, 012040.

\bibitem[Bera and Higgins(1993)]{bera1993arch}
Bera, A.K.; Higgins, M.L. ARCH models: Properties, estimation and testing. \emph{J. Econ. Surv.} \textbf{1993}, \emph{7}, 305--366.

\bibitem[Jiang et~al.(2019)Jiang, Xie, Zhou, and Sornette]{Jiang-Xie-Zhou-Sornette-2019-RPP}
Jiang, Z.Q.; Xie, W.J.; Zhou, W.X.; Sornette, D. Multifractal analysis of financial markets. \emph{Rep. Prog. Phys.} \textbf{2019}, \emph{82}, 125901.

\bibitem[Peng et~al.(1994)Peng, Buldyrev, Havlin, Simons, Stanley, and Goldberger]{peng1994mosaic}
Peng, C.K.; Buldyrev, S.V.; Havlin, S.; Simons, M.; Stanley, H.E.; Goldberger, A.L. Mosaic organization of DNA nucleotides. \emph{Phys. Rev. E} \textbf{1994}, \emph{49}, 1685.

\bibitem[Hurst(1951)]{hurst1951long}
Hurst, H.E. Long-term storage capacity of reservoirs. \emph{Trans. Am. Soc. Civ. Eng.} \textbf{1951}, \emph{116}, 770--799.

\bibitem[Zhou(1996)]{zhou1996high}
Zhou, B. High-frequency data and volatility in foreign-exchange rates. \emph{J. Bus. Econ. Stat.} \textbf{1996}, \emph{14}, 45--52.

\bibitem[Bandi and Russell(2006)]{bandi2006separating}
Bandi, F.M.; Russell, J.R. Separating microstructure noise from volatility. \emph{J. Financ. Econ.} \textbf{2006}, \emph{79}, 655--692.

\bibitem[Hansen and Lunde(2006)]{hansen2006realized}
Hansen, P.R.; Lunde, A. Realized variance and market microstructure noise. \emph{J. Bus. Econ. Stat.} \textbf{2006}, \emph{24}, 127--161.

\bibitem[Liu et~al.(2015)Liu, Patton, and Sheppard]{liu2015does}
Liu, L.Y.; Patton, A.J.; Sheppard, K. Does anything beat 5-minute RV? A comparison of realized measures across multiple asset classes. \emph{J. Econom.} \textbf{2015}, \emph{187}, 293--311.

\bibitem[Andersen et~al.(2000)Andersen, Bollerslev, Diebold, and Labys]{andersen2000exchange}
Andersen, T.G.; Bollerslev, T.; Diebold, F.X.; Labys, P. Exchange rate returns standardized by realized volatility are (nearly) Gaussian. \emph{Multinatl. Financ. J.} \textbf{2000}, \emph{4}, 159--179.

\bibitem[Andersen et~al.(2007)Andersen, Bollerslev, and Dobrev]{andersen2007no}
Andersen, T.G.; Bollerslev, T.; Dobrev, D. No-arbitrage semi-martingale restrictions for continuous-time volatility models subject to leverage effects, jumps and iid noise: Theory and testable distributional implications. \emph{J. Econom.} \textbf{2007}, \emph{138}, 125--180.

\bibitem[Takaishi(2012)]{takaishi2012finite}
Takaishi, T. Finite-sample effects on the standardized returns of the Tokyo Stock Exchange. \emph{Procedia-Soc. Behav. Sci.} \textbf{2012}, \emph{65}, 968--973.

\bibitem[Takaishi(2010)]{takaishi2010analysis}
Takaishi, T. Analysis of realized volatility in superstatistics. \emph{Evol. Institutional Econ. Rev.} \textbf{2010}, \emph{7}, 89--99.

\bibitem[Matteo et~al.(2005)Matteo, Aste, and Dacorogna]{MATTEO2005827}
Matteo, T.D.; Aste, T.; Dacorogna, M.M. Long-term memories of developed and emerging markets: Using the scaling analysis to characterize their stage of development. \emph{J. Bank. Financ.} \textbf{2005}, \emph{29}, 827--851.

\bibitem[Takaishi and Adachi(2020)]{takaishi2019market}
Takaishi, T.; Adachi, T. Market efficiency, liquidity, and multifractality of Bitcoin: A dynamic study. \emph{Asia-Pac. Financ. Mark.} \textbf{2020}, \emph{27}, 145--154.

\bibitem[Zhou(2012)]{zhou2012finite}
Zhou, W.X. Finite-size effect and the components of multifractality in financial volatility. \emph{Chaos Solitons Fractals} \textbf{2012}, \emph{45}, 147--155.

\bibitem[Grech and Pamu{\l}a(2012)]{grech2012multifractal}
Grech, D.; Pamuła, G. Multifractal background noise of monofractal signals. \emph{Acta Phys. Pol. A} \textbf{2012}, \emph{121}, B-34--B-39.

\bibitem[Grech and Pamu{\l}a(2013)]{grech2013multifractal}
Grech, D.; Pamuła, G. On the multifractal effects generated by monofractal signals. \emph{Phys. A Stat. Mech. Its Appl.} \textbf{2013}, \emph{392}, 5845--5864.

\bibitem[Kwapie{\'n} et~al.(2023)Kwapie{\'n}, Blasiak, Dro{\.z}d{\.z}, and O{\'s}wi{\k{e}}cimka]{kwapien2023genuine}
Kwapień, J.; Blasiak, P.; Drożdż, S.; Oświęcimka, P. Genuine multifractality in time series is due to temporal correlations. \emph{Phys. Rev. E} \textbf{2023}, \emph{107}, 034139.

\bibitem[Kluszczy{\'n}ski et~al.(2025)Kluszczy{\'n}ski, Dro{\.z}d{\.z}, Kwapie{\'n}, Stanisz, and W{\k{a}}torek]{kluszczynski2025disentangling}
Kluszczyński, R.; Drożdż, S.; Kwapień, J.; Stanisz, T.; Wątorek, M. Disentangling sources of multifractality in time series. \emph{Mathematics} \textbf{2025}, \emph{13}, 205.

\bibitem[Takaishi(2020)]{takaishi2020rough}
Takaishi, T. Rough volatility of Bitcoin. \emph{Financ. Res. Lett.} \textbf{2020}, \emph{32}, 101379.

\bibitem[Garcin and Grasselli(2022)]{garcin2022long}
Garcin, M.; Grasselli, M. Long versus short time scales: The rough dilemma and beyond. \emph{Decis. Econ. Financ.} \textbf{2022}, \emph{45}, 257--278.

\bibitem[P{\'e}rez-Salinas et~al.(2020)P{\'e}rez-Salinas, Cervera-Lierta, Gil-Fuster, and Latorre]{perez2020data}
Pérez-Salinas, A.; Cervera-Lierta, A.; Gil-Fuster, E.; Latorre, J.I. Data re-uploading for a universal quantum classifier. \emph{Quantum} \textbf{2020}, \emph{4}, 226.

\bibitem[P{\'e}rez-Salinas et~al.(2021)P{\'e}rez-Salinas, L{\'o}pez-N{\'u}{\~n}ez, Garc{\'\i}a-S{\'a}ez, Forn-D{\'\i}az, and Latorre]{perez2021one}
Pérez-Salinas, A.; López-Núñez, D.; García-Sáez, A.; Forn-Díaz, P.; Latorre, J.I. One qubit as a universal approximant. \emph{Phys. Rev. A} \textbf{2021}, \emph{104}, 012405.

\bibitem[Ding and Granger(1996)]{ding1996modeling}
Ding, Z.; Granger, C.W. Modeling volatility persistence of speculative returns: A new approach. \emph{J. Econom.} \textbf{1996}, \emph{73}, 185--215.

\bibitem[Haas(2009)]{haas2009persistence}
Haas, M. Persistence in volatility, conditional kurtosis, and the Taylor property in absolute value GARCH processes. \emph{Stat. Probab. Lett.} \textbf{2009}, \emph{79}, 1674--1683.

\bibitem[Takaishi and Adachi(2018)]{takaishi2018taylor}
Takaishi, T.; Adachi, T. Taylor effect in Bitcoin time series. \emph{Econ. Lett.} \textbf{2018}, \emph{172}, 5--7.

\bibitem[Bouchaud et~al.(2001)Bouchaud, Matacz, and Potters]{bouchaud2001leverage}
Bouchaud, J.P.; Matacz, A.; Potters, M. Leverage effect in financial markets: The retarded volatility model. \emph{Phys. Rev. Lett.} \textbf{2001}, \emph{87}, 228701.

\bibitem[Qiu et~al.(2006)Qiu, Zheng, Ren, and Trimper]{qiu2006return}
Qiu, T.; Zheng, B.; Ren, F.; Trimper, S. Return-volatility correlation in financial dynamics. \emph{Phys. Rev. E} \textbf{2006}, \emph{73}, 065103.

\bibitem[Shen and Zheng(2009)]{shen2009return}
Shen, J.; Zheng, B. On return-volatility correlation in financial dynamics. \emph{Europhys. Lett.} \textbf{2009}, \emph{88}, 28003.

\bibitem[Zumbach(2003)]{zumbach2003volatility}
Zumbach, G. Volatility processes and volatility forecast with longmemory. \emph{Quant. Financ.} \textbf{2003}, \emph{4}, 70.
\clearpage 
\bibitem[Zumbach(2009)]{zumbach2009time}
Zumbach, G. Time reversal invariance in finance. \emph{Quant. Financ.} \textbf{2009}, \emph{9}, 505--515.

\bibitem[Gopikrishnan et~al.(1998)Gopikrishnan, Meyer, Amaral, and Stanley]{gopikrishnan1998inverse}
Gopikrishnan, P.; Meyer, M.; Amaral, L.N.; Stanley, H.E. Inverse cubic law for the distribution of stock price variations. \emph{Eur. Phys. J. B-Condens. Matter Complex Syst.} \textbf{1998}, \emph{3}, 139--140.

\end{thebibliography}





\isPreprints{}{
\end{adjustwidth}
} 
\end{document}